\documentclass[
reprint,
longbibliography,
amsmath, 
amssymb, 
aps, 
superscriptaddress,
pra,
]
{revtex4-2}

\usepackage{graphicx}
\usepackage{dcolumn}
\usepackage{bm}
\usepackage{braket} 
\usepackage{hyperref}
\usepackage{subfigure} 
\usepackage{nccmath}
\usepackage{amsthm} 
\usepackage{mathrsfs} 
\usepackage{xfrac} 
\usepackage{pifont}
\usepackage[export]{adjustbox}
\usepackage{bbm} 
\usepackage[utf8]{inputenc} 
\usepackage[normalem]{ulem}
\usepackage[usenames,dvipsnames,svgnames,table]{xcolor}
\usepackage[sort&compress]{natbib}
\usepackage{color}
\usepackage{longtable}
 
\usepackage{commath}
\usepackage{multirow}
\usepackage{indentfirst}

\begin{document}
\title{Unbalanced penalization: A new approach to encode inequality constraints of combinatorial problems for quantum optimization algorithms}

\author{J. A. Monta\~nez-Barrera}
 	\altaffiliation{Corresponding author: J. A. Monta\~nez-Barrera; j.montanez-barrera@fz-juelich.de}
	\affiliation{Jülich Supercomputing Centre, Institute for Advanced Simulation, Forschungszentrum Jülich, 52425 Jülich, Germany}
\author{Dennis Willsch}
	\affiliation{Jülich Supercomputing Centre, Institute for Advanced Simulation, Forschungszentrum Jülich, 52425 Jülich, Germany}
\author{A. Maldonado-Romo}
	\affiliation{Centro de investigaci\'on en Computaci\'on, Instituto Polit\'ecnico Nacional, M\'exico}
	\affiliation{Entropica Labs, 186b Telok Ayer Street, Singapore 068632}
\author{Kristel Michielsen}
	\affiliation{Jülich Supercomputing Centre, Institute for Advanced Simulation, Forschungszentrum Jülich, 52425 Jülich, Germany}
	\affiliation{AIDAS, 52425 Jülich, Germany}
	\affiliation{RWTH Aachen University, 52056 Aachen, Germany}
\begin{abstract}

Solving combinatorial optimization problems of the kind that can be codified by quadratic unconstrained binary optimization (QUBO) is a promising application of quantum computation. Some problems of this class suitable for practical applications such as the traveling salesman problem (TSP), the bin packing problem (BPP), or the knapsack problem (KP) have inequality constraints that require a particular cost function encoding. The common approach is the use of slack variables to represent the inequality constraints in the cost function. However, the use of slack variables considerably increases the number of qubits and operations required to solve these problems using quantum devices. In this work, we present an alternative method that does not require extra slack variables and consists of using an unbalanced penalization function to represent the inequality constraints in the QUBO. This function is characterized by larger penalization when the inequality constraint is not achieved than when it is. We evaluate our approach on the TSP, BPP, and KP, successfully encoding the optimal solution of the original optimization problem near the ground state cost Hamiltonian. Additionally, we employ D-Wave Advantage and D-Wave hybrid solvers to solve the BPP, surpassing the performance of the slack variables approach by achieving solutions for up to 29 items, whereas the slack variables approach only handles up to 11 items. This new approach can be used to solve combinatorial problems with inequality constraints with a reduced number of resources compared to the slack variables approach using quantum annealing or variational quantum algorithms.

\begin{description}
	\vspace{0.2cm}
	\item[Keywords] QUBO; inequality constraints; knapsack; bin packing; traveling salesman problem; quantum optimization; VQA; QAOA; quantum annealing; D-Wave Advantage; D-Wave Hybrid; combinatorial optimization.
\end{description}

\end{abstract}

\maketitle

\section{\label{int}Introduction}

The exploration of quantum enhancement in combinatorial optimization problems has been extended over several years. There are three main reasons for this; first, many combinatorial optimization problems can be encoded in Hamiltonians, the ground state being the optimal solution of the problem \cite{Lucas2014, Kochenberger2014}; second, they are commonly hard to solve and have practical applications \cite{Ohzeki2020}; and third, the quantum algorithms to solve these problems need few resources and can be tested on current state-of-the-art quantum hardware \cite{Harrigan2021, Niroula2022}.

The usual approach for encoding combinatorial optimization problems on quantum processing units (QPUs) is to transform them in their quadratic unconstrained binary optimization (QUBO) representation and obtain the cost Hamiltonian after a change of variables. In the QUBO encoding, the constraints of the combinatorial optimization problems are added to the cost function as penalization terms along with the objective function.

However, we are currently in a noisy intermediate-scale quantum (NISQ) stage and QPUs are restricted in terms of qubits number, noise level, and circuit depth \cite{Preskill2018}. The main consequence is that numerous problems solvable by quantum algorithms such as Shor's algorithm \cite{shor1994factoring,shor1997algorithm} remain unpractical to test on real quantum hardware. Despite this fact, solving combinatorial optimization problems is possible on current hardware. For example, on gate-based QPUs, Variational Quantum Algorithms (VQA) \cite{Cerezo2021} are applicable even with the above limitations thanks to their short depth and resistance to noise \cite{Khairy2020}. Besides, quantum annealers, using the quantum annealing (QA) principle~\cite{Apolloni89,Finnila1994QuantumAnnealing, KadowakiNishimori1998QuantumAnnealing}, are another quantum technology used to solve combinatorial optimization problems \cite{Falco2011, Ayanzadeh2021, Willsch2021BenchmarkAdvantage}. In this respect, QA shows some advantages under certain conditions compared to classical annealing \cite{Heim2015, Yan2022, Tasseff2022}.

Of the aforementioned VQA algorithms, the most studied for combinatorial optimization problems is the Quantum Approximate Optimization algorithm (QAOA) \citep{Farhi2014}. Even if it is still unclear whether QAOA will give any advantage compared with other algorithms~\cite{Willsch2019BenchmarkingQAOA}, its simplicity makes it of great interest for analytical and practical purposes \cite{Harrigan2021}. In the simplest QAOA version, the cost Hamiltonian of a combinatorial optimization problem is encoded in a parametric unitary gate along with a ``mixer", a second parametric unitary gate that does not commute with the first unitary gate. In this context, the parameters are adjusted to minimize the expectation value of the cost Hamiltonian using a classical optimizer. Multiple approaches for solving combinatorial optimization problems using QAOA or QA can be found in the literature, for example, in logistics \cite{Jiang2022}, finance \cite{Orus2019, Niroula2022, Souza2022, Mugel2022}, energy \cite{Sharabiani2021}, communications \cite{Urgelles2022}, automotive industry \cite{Luckow2021}, traffic signaling \cite{Inoue2021}, among others.

From the different sets of problems that can be solved using quantum hardware, the ones with inequality constraints require an extra number of variables to get their QUBO representation. For example, the BPP and the TSP have many inequality constraints that increase with the number of items and cities, respectively. The usual approach implemented in software development kits (SDKs) such as Qiskit~\cite{Qiskit} or D-Wave Ocean~\cite{DWOceanSDK} is to use slack variables to encode the inequality constraints \cite{Glover2019}. Still, such an approach increments the number of qubits and connections needed to solve these problems. We remark that an alternative approach to handle inequality constraints with penalization would be to develop direct algorithms that respect the constraints by construction, as illustrated in \cite{Glover2022} for the asset exchange problem. 

In this paper, we propose an alternative method to encode inequality constraints in the QUBO formulation of combinatorial optimization problems. In this new heuristic encoding method, inequality constraints are encoded using an unbalanced penalization formula. This formulation adds larger penalization when the inequality constraint is not fulfilled than when it is. We test our method on the TSP, the BPP, and the KP using OpenQAOA \cite{Sharma2022}, a python-based library developed by Entropica Labs, and the J\"ulich universal quantum computer simulator (JUQCS) \cite{DeRaedt2007MassivelyParallel,DeRaedt2018MassivelyParallel,Willsch2021JUQCSGQAOA} for problems with up to 43 qubits. In the majority of the cases, the optimal solution of the original combinatorial optimization problem is encoded in the ground state of the cost Hamiltonian of their respective QUBO. However, in some cases, the optimal solution is in the vicinity of the ground state. For such cases, our method still provides a good approximation given the whole set of eigenvalues.

 Additionally, we test the performance for finding solutions for the BPP using the D-Wave Hybrid solver and the quantum annealer D-Wave Advantage 5.3 System JUPSI, a quantum computing system developed by D-Wave Systems Inc that has more than 5000 qubits located in J\"ulich, Germany. Our findings demonstrate that our method outperforms existing approach in terms of both the quantity and quality of solutions obtained for the BPP. Specifically, using the unbalanced penalization approach, we achieve remarkable results on both systems. With the D-Wave Advantage system, we obtain solutions for up to 7 nodes, while with the D-Wave Hybrid solver, we achieve solutions for up to 29 nodes. In contrast, the traditional slack variables approach only handles solutions for up to 7 and 11 nodes for the D-Wave Advantage and D-Wave Hybrid, respectively. 

The rest of the paper is organized as follows. Sec. \ref{Sec:QUBO} provides a description of the implementation of the combinatorial optimization problems using QUBOs and an overview of the TSP, BPP, and KP. Section \ref{NewApproach} presents a description of the {\it unbalanced penalization} approach and a new metric called the coefficient of performance (CoP) to study the efficiency of QAOA to solve combinatorial optimization problems. Section \ref{Sec:Results} presents the results and discussion. Finally, Section \ref{Sec:Conclusion} provides some conclusions. The source code for the new method setup and implementation on the TSP, BPP, and KP can be found at \url{https://jugit.fz-juelich.de/qip/unbalanced-penalizations-qubo}.

\section{Quadratic unconstrained binary optimization} \label{Sec:QUBO}
The set of combinatorial problems that can be represented by the QUBO formulation are characterized by functions of the form

\begin{equation}
f(\mathrm{x}) = \frac{1}{2}\sum_{i=1}^{n} \sum_{j=1}^n q_{ij} x_{i} x_{j}, 
\end{equation}
where $n$ is the number of variables, $q_{ij} \in \mathbb{R}$ are coefficients associated to the specific problem, and $x_i \in \{0,1\}$ are the binary variables of the problem. Note that $x_{i} x_{i} \equiv x_{i}$ and $q_{ij} = q_{ji}$ in this formulation. Therefore, the general form of a combinatorial optimization problem solvable by QPUs is given by the cost function

\begin{equation}\label{QUBO_form}
f(\mathrm{x}) = \sum_{i=1}^{n-1} \sum_{j > i}^n q_{ij}x_{i}x_{j} + \sum_{i=1}^n q_{ii} x_i,
\end{equation}
and equality constraints are given by

\begin{equation}
\sum_{i=1}^n c_i x_i = C, \ c_i \in \mathbb{Z},
\end{equation}
and inequality constraints are given by

\begin{equation}\label{inequality}
\sum_{i=1}^n l_i x_i \le B, \ l_i \in \mathbb{Z}.
\end{equation}

To transform these problems into the QUBO formulation the constraints are added as penalization terms. In this respect, the equality constraints are included in the cost function using the following penalization term

\begin{equation}\label{EQ_F}
\lambda_0 \left(\sum_{i=1}^n c_i x_i - C\right)^2,
\end{equation}
where $\lambda_0$ is a penalization coefficient that should be chosen to guarantee that the equality constraint is fulfilled and $C$ is a constant value. In the case of inequality constraint, the common approach is to use a slack variable \cite{Glover2019, Verma2022}. The slack variable, $S$, is an auxiliary variable that makes a penalization term vanish when the inequality constraint is achieved.

 \begin{equation}\label{ineq}
 B - \sum_{i=1}^n l_i x_i - S = 0.
 \end{equation}
 Therefore, when Eq.(\ref{inequality}) is satisfied, Eq.(\ref{ineq}) is already zero. This means the slack variable, $S$, must be in the range $0 \le S \le B - \min_x \sum_{i=1}^n l_i x_i$. To represent the $slack$ variable in binary form, the slack is decomposed in $N = \lfloor \log_2(\max_x B - \sum_{i=1}^n l_i x_i) + 1\rfloor $ binary variables: 

\begin{equation}\label{SB} 
S = \sum_{k=0}^{N-1} 2^k s_k,
\end{equation}
where $s_k$ are the slack binary variables. Then, the inequality constraints are added as penalization terms by 
 
 \begin{equation}\label{Ineq_EF}
 \lambda_1  \left(B - \sum_{i=1}^n l_i x_i - \sum_{k=0}^{N-1} 2^k s_k\right)^2.
 \end{equation}
 
Combining Eq.(\ref{QUBO_form}) and the two kinds of constraints Eq.(\ref{EQ_F}) and Eq.(\ref{Ineq_EF}), the general QUBO representation of a given combinatorial optimization problem is given by

\begin{widetext}
 \begin{equation}\label{QUBO}
 \min_x \left(\sum_{i=1}^{n-1} \sum_{j > i}^nc_{ij}x_{i}x_{j} + \sum_{i=1}^n h_i x_i + \lambda_0 \left(\sum_{i=1}^n q_i x_i - C\right)^2
+  \lambda_1  \left(B - \sum_{i=1}^n l_i x_i - \sum_{k=0}^{N-1} 2^k s_k\right)^2\right).
 \end{equation}
\end{widetext}

Following the same principle, more constraints can be added and note that after some manipulations, Eq.(\ref{QUBO}) can be rewritten in the form of Eq.(\ref{QUBO_form}). The last step to represent the QUBO problem on QPUs is to change the $x_i$ variables to spin variables $z_i \in \{1, -1\}$ by the transformation $x_i = (1 - z_i) / 2$. Hence, Eq.(\ref{QUBO_form}) represented in terms of the cost Hamiltonian model reads

\begin{equation}\label{IsingH}
H_c(\mathrm{z}) = \sum_{i=1}^{n-1}\sum_{j>i}^n q_{ij} (1 - z_i) (1 - z_j)/4 + \sum_{i=1}^n q_{ii} (1 - z_i)/2.
\end{equation} 

\subsection{Traveling salesman problem}
In the TSP, a traveler has to stop by a set of cities, finding the shortest route to visit every city once and return to the starting point. This problem is normally solved classically for thousands to millions of variables using heuristic techniques \cite{Helsgaun2006, Applegate2006}. For instance, the Concorde TSP solver \cite{concorde-website} can solve all 110 problems from TSPLIB \cite{TSPLIB} to optimality, the largest having 85000 cities.

\begin{figure*}[!tbh]
\centering
\includegraphics[width=18cm]{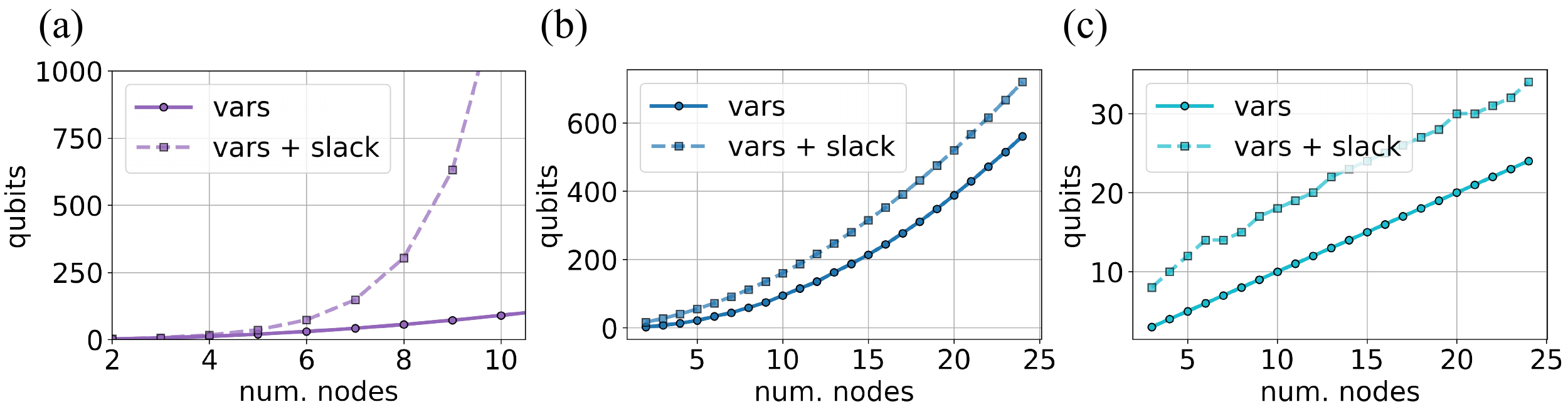}
\caption{\label{variables_all} Number of qubits needed to solve (a) the TSP for different numbers of cities, (b) the BPP for different numbers of items, and (c) the KP for different numbers of items. The solid line represents the number of variables of the problem and the dashed line represents the variables of the problem plus the slack variables needed to represent the inequality constraints of the problem for a different number of nodes.}
\end{figure*}

Different formulations of the TSP exist, but in this work, we focus on the Dantzig-Fulkerson-Johnson (DFJ) formulation \cite{Grotschel2008}, which is the linear programming (LP) version of the TSP. Note that while there are other formulations of this problem that do not require inequality constraints, e.g., \cite{Lucas2014}, the DFJ formulation is used here to test the capabilities of the unbalanced penalization method to encode multiple inequality constraints. In this formulation, the problem is given by

\begin{equation}\label{TSP}
\min\sum_{i=1}^{n}\sum_{j \neq i, j=1}^{n} c_{ij}x_{ij},
\end{equation}
subject to the set of constraints,

\begin{equation}\label{TSPEC1}
\sum_{i=1, i \neq j}^{n} x_{ij} = 1 \qquad \forall j = 1, ..., n,
\end{equation}

\begin{equation}\label{TSPEC2}
\sum_{j=1, j \neq i}^{n} x_{ij} = 1 \qquad \forall i = 1, ..., n,
\end{equation}

\begin{equation}\label{TSPIC}
\sum_{i \in Q} \sum_{j \neq i, j \in Q} x_{ij}\leq |Q| - 1 \ \  \forall Q \subsetneq \{1, ..., n\}, |Q| \ge 2,
\end{equation}

\begin{equation}\label{BPPB1}
x_{ij}\in  \{0,1\} \qquad \forall i=1,..,n \qquad \forall j=1,..,n,
\end{equation}
where $n$ is the number of cities (nodes), $c_{ij}$ is the distance between the cities $i$ and $j$, $x_{ij}$ is a binary variable that represents if the path from the city $i$ to the city $j$ is used or not, and $Q$ represents a sub-tour on a set of cities. From the above equations, Eq.(\ref{TSP}) is the cost function for the distance traveled, Eq.(\ref{TSPEC1}) and Eq.(\ref{TSPEC2}) restrict the traveler to take only one path to enter and one to leave a city, and Eq.(\ref{TSPIC}) are the inequality constraints that prohibit sub-tours in the solution. Of all the inequality constraints analyzed in this work, constraint Eq.(\ref{TSPIC}) is the most expensive in terms of the number of slack variables. Fig.~\ref{variables_all} (a) shows the number of qubits needed to represent the TSP for a different number of cities. 

For this work, we randomly place $n$ cities on a 50x50 grid with $c_{ij}$ being the Euclidian distance between cities. We vary from 2 to 7 cities and for problems up to 5 cities, we generate 10 random problems to test the generalization of our method. 

\subsection{Knapsack problem}

In the KP, a set of items with associated weights and values should be stored in a knapsack. The problem is to maximize the value of the items (nodes) transported in the knapsack. The KP is restricted by the maximum weight the knapsack can carry. Typically, this problem is solved classically for thousands of variables and accepts pseudo-polynomial algorithms, i.e., an algorithm with complexity  bounded by a polynomial in the number of items and the maximum weight \cite{Martello1990}. 

KP is the simplest nontrivial integer programming model with binary variables, only one constraint, and positive coefficients. It is formally defined by

\begin{equation}
\max  \sum_{i=1}^{n} p_{i} x_{i},
\end{equation}

\begin{equation}
\sum_{i=1}^{n} w_{i} x_{i} \leq W, 
\end{equation}
where $n$ is the number of items, $p_{i}$ and $w_{i}$ are the value and weight of the $ith$ item, respectively, $x_i$ is the binary variable that represents whether the $ith$ item is in the knapsack or not, and W is the maximum weight that the knapsack can transport. Fig.~\ref{variables_all} (c) shows the number of qubits needed to solve the KP for a different number of items. The set of problems is created with weights and values selected randomly ranging between 1 and 63 for the values, between 1 and 127 for the weights, and the knapsack maximum weight is equal to 70\% of the sum of the weights of all the items.

\subsection{Bin packing problem}\label{BPP}

In the BPP, a collection of items must be stored in the minimum possible number of bins. In this case, the items have a weight associated and the bins are restricted to carry a maximum weight. Commonly, this problem accepts hundreds of bins, but it has been shown to be NP-Hard in the strongest sense, i.e., there is no pseudo-polynomial algorithm to solve this problem \cite{Martello1990}. The problem has many real-world applications such as loading trucks with a weight restriction \cite{Hessler2020}, container scheduling \cite{Yan2022Book}, or design of FPGA chips \cite{Kroes2020}. Its formulation is given by

\begin{equation}\label{BPPmin}
\min \sum_{j=1}^m y_j,
\end{equation}
subject to:

\begin{equation}\label{BPPIC}
\sum_{i=1}^n w_i x_{ij} \le B y_j \qquad  \forall j=1,...,m,
\end{equation}

\begin{equation} \label{BPPEC}
\sum_{j=1}^m x_{ij} = 1  \qquad \forall i = 1, ..., n,
\end{equation}

\begin{equation}\label{BPPB1}
x_{ij}\in  \{0,1\} \qquad \forall i=1,..,n \qquad \forall j=1,..,m,
\end{equation}
\begin{equation}\label{BPPB2}
y_{j}\in  \{0,1\} \qquad \forall j=1,..,m, 
\end{equation}
where $n$ is the number of items (nodes), $m$ is the number of bins, $w_{i}$ is the {\it i-th} item weight, $B$ is the maximum weight of each bin, $x_{ij}$ and $y_j$ are binary variables that represent if the item $i$ is in the bin $j$, and whether bin $j$ is used or not, respectively. From the above equations, Eq.(\ref{BPPmin}) is the cost function to minimize the number of bins, Eq.(\ref{BPPIC}) is the inequality constraint for the maximum weight of a bin, Eq.(\ref{BPPEC}) is the equality constraint to restrict that an item is only in one of the bins, and Eqs.(\ref{BPPB1}) and (\ref{BPPB2}) means that $y_i$ and $x_{ij}$ are binary variables.

Fig.~\ref{variables_all} (b) shows the number of variables needed to solve the BPP when the number of bins equals the number of items. The weights are chosen randomly with values between 4 and 20 and the maximum bin weight is equal to 20. Without losing generality, two further simplifications are made, the first assigns the first item to the first bin $x_{11} = 1$ and therefore $x_{1j} = 0 \ \forall \ j \in \{2, ..., m\}$. Second replaces $y_j = 1 \ \forall \ j \in \ \{ 1, 2, ..., N^{bin}_{min}\}$ with $N^{bin}_{min} = \lceil (\sum_{i=1}^n s(i)\rceil) / B$ because the minimum of bins required is known.

\begin{figure}[!tbh]
\centering
\includegraphics[width=8.5cm]{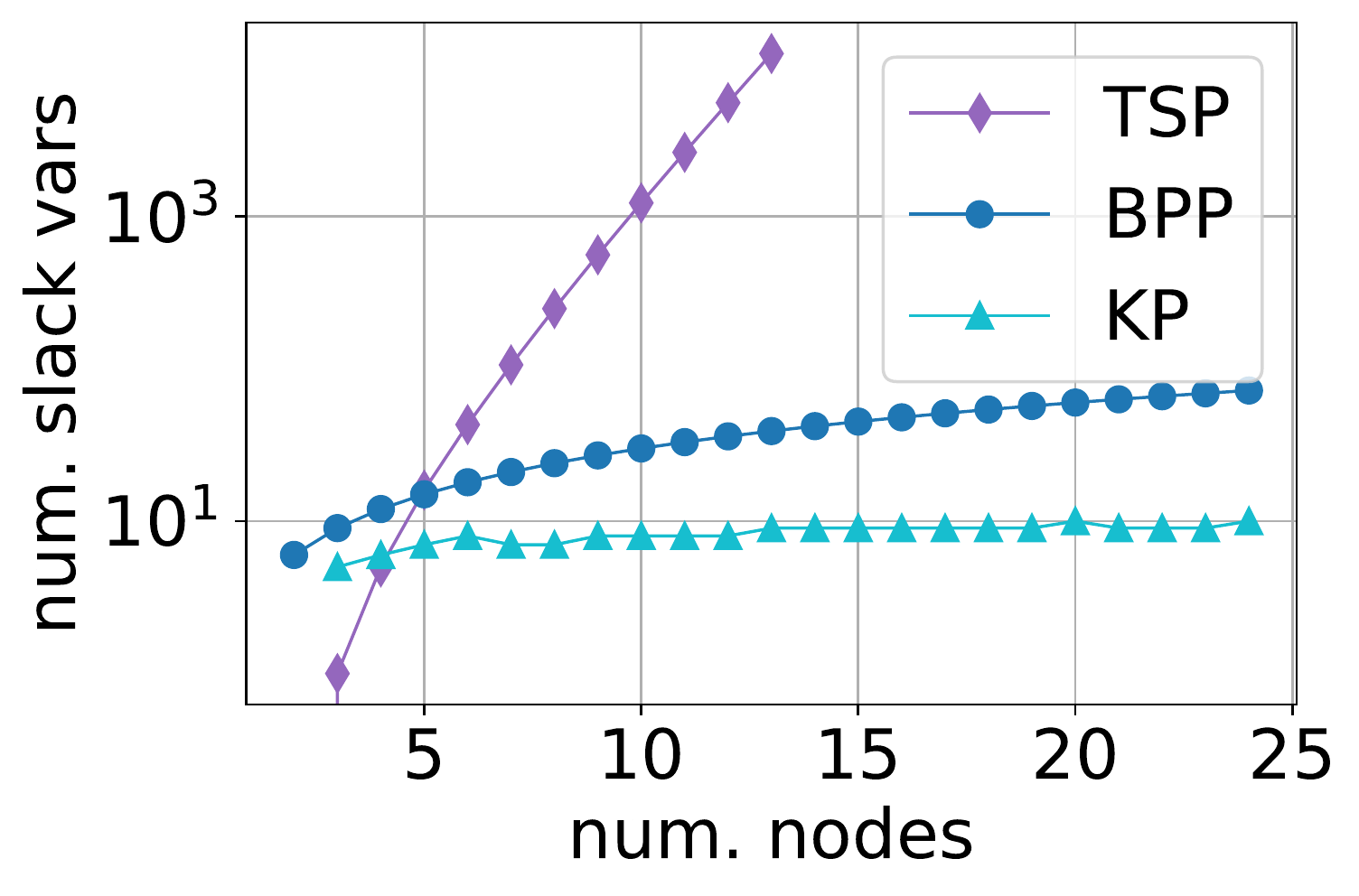}
\caption{\label{slack} The number of slack variables needed to solve the TSP (diamonds), the BPP (circles), and the KP (triangles) based on the number of nodes. The solid lines are guides to the eye.}
\end{figure}

An estimate of the number of slack variables needed to represent the QUBO of the TSP, the BPP, and the KP is shown in Fig.~\ref{slack}. In the TSP, the number of slack variables increases exponentially with the number of cities added. The BPP requires many slack binary variables increasing proportionally to the number of bins of the problem. Lastly, in the KP, the number of slack variables needed is constant and depends on the maximum weight allowed in the knapsack.

\section{Unbalanced penalization}\label{NewApproach}
\begin{figure}[!tbh]
\centering
\includegraphics[width=8.5cm]{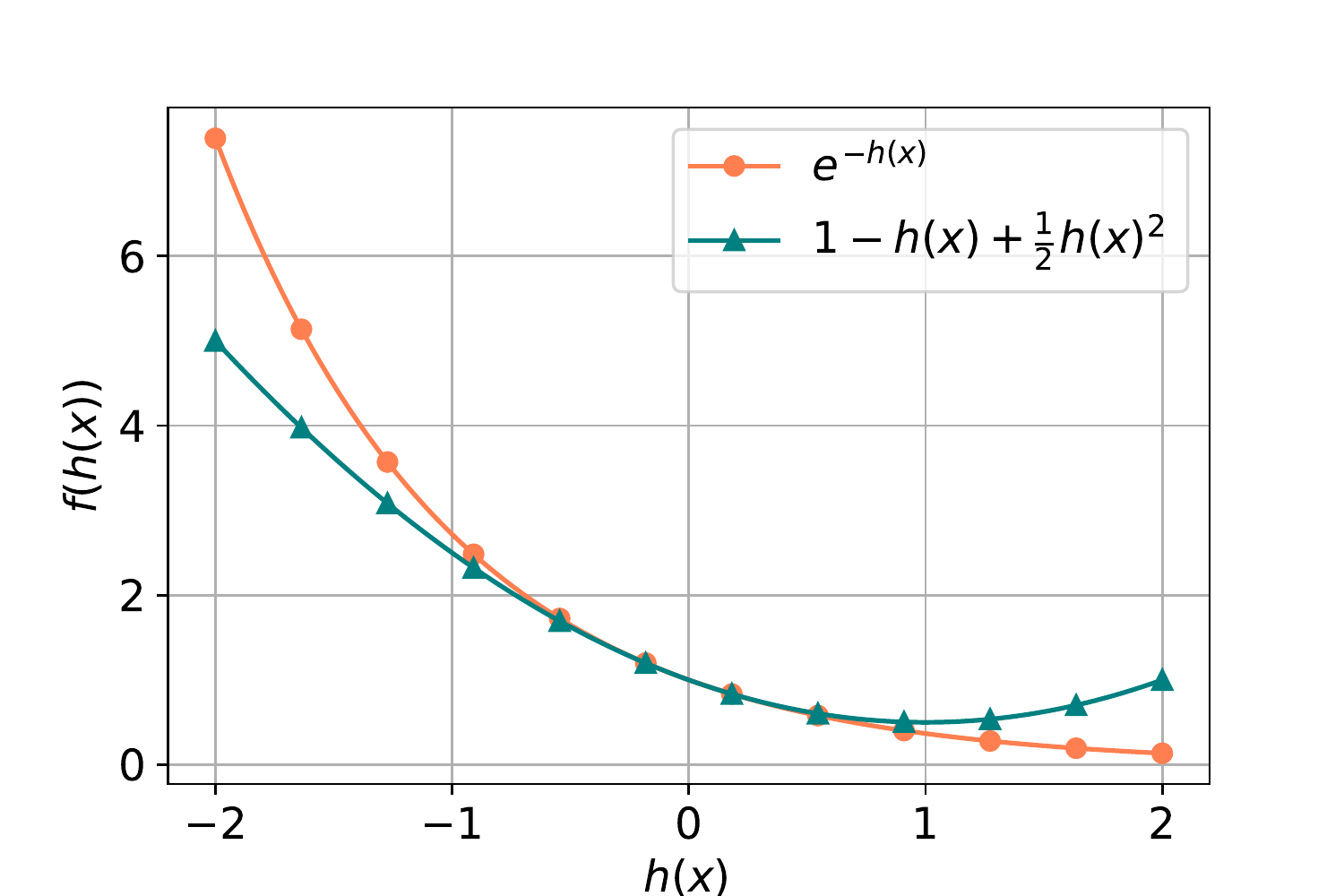}
\caption{\label{unbalanced_eq} Comparison between $e^{-h(\mathrm{x})}$ and the unbalanced function $1 - h(\mathrm{x}) + \frac{1}{2} h(\mathrm{x})^ 2$.}
\end{figure}

The unbalanced penalization approach we propose is an approximation for including inequality constraints of combinatorial optimization problems in the QUBO representation of the cost function. This method does not require additional slack variables, and the inequality constraint of Eq.~(\ref{inequality}) given by

\begin{equation}\label{Eq:ineq}
h(\mathrm{x}) =  B - \sum_i l_i x_i \ge 0,
\end{equation}
is replaced by a penalization term in the QUBO formulation that makes larger penalization for negative terms and smaller for positive ones. The idea of the unbalanced penalization function comes from the shape of the exponential decay curve, $f(x) = e^{-h(\mathrm{x})}$. In this function, positive values of $h(\mathrm{x})$, given by Eq.~(\ref{Eq:ineq}), make $f(\mathrm{x}) \approx 0$ while negative values make it grow exponentially, as shown in Fig.~\ref{unbalanced_eq}. However, the exponential function cannot be encoded as a QUBO penalization term. Therefore, we consider an expansion of the exponential function to quadratic order. The penalization term is given by

\begin{equation}\label{exp_fun_s}
e^{-h(\mathrm{x})} \approx 1 - h(\mathrm{x}) + \frac{1}{2} h(\mathrm{x})^ 2.
\end{equation}

In general, we modify Eq.~(\ref{exp_fun_s}) to include free parameters to be adjusted for the different kinds of problems. Therefore, Eq.~(\ref{exp_fun_s}) is rewritten by

\begin{align}
\zeta(x) & = - \lambda_1 h(x) + \lambda_2 h(x)^2 \\
& =  - \lambda_1\left(B - \sum_i l_i x_i\right) + \lambda_2 \left(B - \sum_i l_i x_i\right)^2,
\end{align}
where $\lambda_{1,2}$ are a set of multipliers that can be tuned for the specific problem and the constant term is removed because the position of the cost function's minimum is independent of the constant term. Note that this approach can be extended to $g(\mathrm{x}) =  B - \sum_i l_i x_i \le 0$, in that case $f(x) = e^{g(x)}$. The new cost function based on Eq.~(\ref{QUBO}) and the unbalanced penalization approach is given by 

\begin{widetext}
   \begin{equation}\label{QUBO_unbalanced}
 \min_x \left(\sum_{i=1}^n \sum_{j\ne i}^nc_{ij}x_{i}x_{j} + \sum_{i=1}^n h_i x_i + \lambda_0 \left(\sum_{i=1}^n q_i x_i - C\right)^2
- \lambda_1\left(B - \sum_i l_i x_i\right)+  \lambda_2 \left(B - \sum_i l_i x_i\right)^2 \right),
 \end{equation}
\end{widetext}

To tune the $\lambda_{0,1,2}$ parameters of Eq.~(\ref{QUBO_unbalanced}) for the TSP, the BPP, and the KP, we use the Nelder-Mead optimization method to explore the set of $\lambda_{0,1,2}$ values that brings the optimal solution of the original combinatorial optimization problems as close as possible to the ground state of the cost Hamiltonian, Eq.~(\ref{IsingH}), of their QUBO representation. For the TSP we use a random problem with 4 cities, for the BPP we use a random problem with 4 items, and for the KP we use a problem with 13 items. The set of values $\lambda_{0,1,2}$ found by the Nelder-Mead optimization method for each problem is presented in Table \ref{Table1}. After we find the best set of values for the specific problem, we test if those values generalize to other random cases. In the random cases used, we find that a set of $\lambda_{0,1,2}$ of one problem works well for different and/or larger problems. Therefore, we use a unique set of these parameters for each problem. However, this is not necessarily the case for other instances of the same problem with a different distribution of parameters. For example, the TSP has some instances that are harder to solve \cite{Cheeseman1991}. In general, one can repeat the procedure mentioned above with multiple cases of the specific distribution to estimate a good set of $\lambda_{0,1,2}$ parameters on the new distribution. The results on these aspects of the unbalanced penalization method are presented in Sec.\ref{eigenvalues_encoding}. 

\begin{table}[ht]
 \caption{\label{Table1}Parameters $\lambda_{0,1,2}$ for the TSP, BPP, and KP used to translate the combinatorial optimization problems into the QUBO representation using the unbalanced penalization approach (Eq.~(\ref{QUBO_unbalanced}).) For all the equality constraints of each problem, we use the same $\lambda_0$ and for the inequality constraints the same $\lambda_{1,2}$.}
\begin{center}
\begin{tabular}{|p{0.1\textwidth}|p{0.1\textwidth}|p{0.1\textwidth}|p{0.1\textwidth}|}\hline
& $\lambda_0$& $\lambda_1$ & $\lambda_2$ \\ 
 \hline
TSP & 38.2584 & 18.2838 & 57.0375\\
BPP & 20.5198 & 7.2949 & 0.8583\\
KP & - & 0.9603 & 0.0371\\
 \hline
\end{tabular}
\end{center}
\end{table}

\subsection*{Quantum Annealing}
Quantum annealing is a computational technique designed to tackle optimization problems by leveraging quantum mechanical effects in the search for low-energy states of Ising Hamiltonians \cite{Kadowaki1998, Johnson2011}. By employing this method, the system can explore a broader solution space and potentially find good-quality solutions more efficiently than classical optimization approaches.

In order to validate the effectiveness of our proposed method on real hardware, we conducted experiments using the D-Wave Advantage and the D-Wave Hybrid Quantum-Classical solver to obtain results for the BPP. To ensure a fair comparison, we selected random BPP problems with the same settings as described in Section \ref{BPP}, without any simplifications.

For the experiments using the D-Wave Advantage, we conducted 20 trials on random problem instances ranging from 3 to 8 nodes (items). Each trial consisted of 5000 samples. For the hybrid solver, we performed 20 experiments with a single run, allowing a maximum execution time of 3 seconds. The problem instances for the hybrid solver ranged from 5 to 31 nodes (items).

In both cases, we utilized the $\lambda_{0,1,2}$ parameters as presented in Table \ref{Table1}. It is important to note that our objective was not to fine-tune the hyperparameters for either solver but rather to demonstrate that under similar conditions, unbalanced penalization yields significantly better solutions in terms of both quantity and quality. Results are presented in Sec. \ref{QA}.

\subsection*{Quantum Approximate Optimization Algorithm}

\begin{figure}[!tbh]
\centering
\includegraphics[width=8cm]{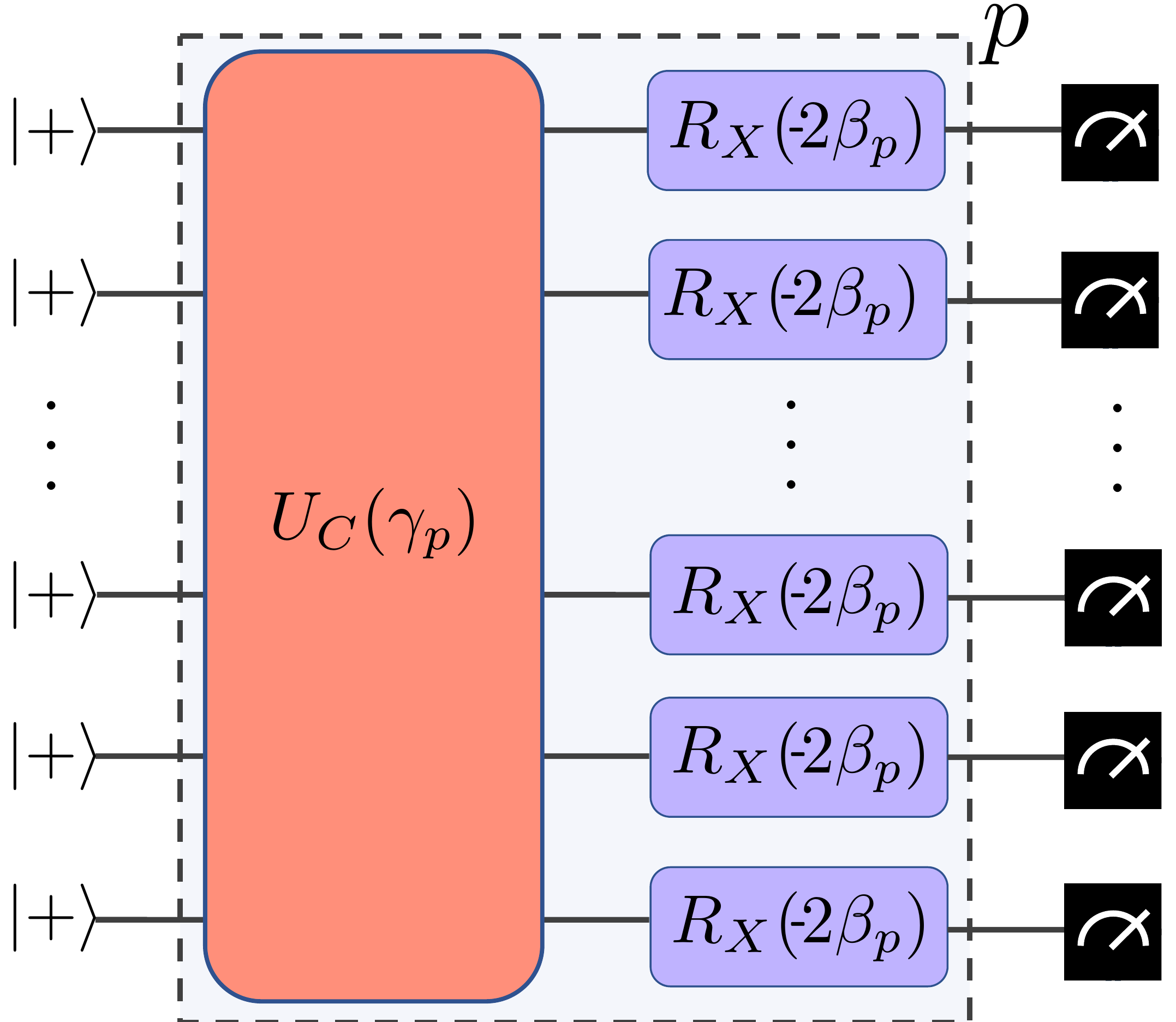}
\caption{\label{cir_qaoa}Schematic representation of QAOA for $p$ layers. The parameters $\gamma$ and $\beta$ for each layer are the ones to be optimized. QAOA is used to compare the slack variables method vs. the unbalanced penalization method for the encoding of the inequality constraints.}
\end{figure}

To test the difference between the slack variables approach vs. the unbalanced penalization encoding on different random problems of the TSP, the BPP, and the KP, we use the QAOA with one layer. It is important to acknowledge that, while the impact of penalization terms of the constraints remains uncertain in terms of their influence on the performance of the Quantum Approximate Optimization Algorithm (QAOA)\cite{Wang2019}, we employ QAOA to demonstrate the contrasting probability distributions resulting from the two approaches. In this case, the cost Hamiltonian, $H_c$, obtained from the QUBO formulation, is translated into a parametric unitary gate given by

\begin{equation}\label{UC}
    U(H_c, \gamma)=e^{-i \gamma H_c},
\end{equation}
 where $\gamma$ is a parameter to be optimized. A second unitary operator applied is 

\begin{equation}\label{UB}
    U(B, \beta)=e^{i \beta X},
\end{equation}
where $\beta$ is the second parameter that must be optimized and $X = \sum_{i=1}^n \sigma_i^x$ with $\sigma_i^x$ the Pauli-x quantum gate applied to qubit $i$. The general QAOA circuit is shown in Fig.~\ref{cir_qaoa}. Here, $R_X(\theta) = e^{-i \frac{\theta}{2} \sigma_x}$, $p$ represents the number of repetitions of the unitary gates Eqs.~\ref{UC} and \ref{UB} with each repetition having separate values for $\gamma_p$ and $\beta_p$, and the initial state is a superposition state $| + \rangle^{\otimes n}$. The case of QAOA with $p=1$ is suitable for visualization of the landscape and in Sec. \ref{QAOAlandscape}, we show a comparison between both encodings and even though they are similar, the addition of extra qubits for the slack variables approach has a great impact on the probability of finding the optimal solution of the original combinatorial optimization problem.

\subsection*{Coefficient of performance}
Additionally, we introduce the coefficient of performance (CoP), a quantifier that expresses how good a quantum algorithm is to give a specific solution compared to random guessing over a set of solutions. In this perspective, we use the CoP to determine how likely it is to find the optimal solution for the TSP, the BPP, and the KP on the local minima of the QAOA algorithm for $p = 1$. The CoP is given by

\begin{equation} 
CoP = \frac{P(^*\mathrm{x})}{P_R},
\end{equation}
where $P(^*\mathrm{x})$ is the probability of getting a bit string $^*\mathrm{x}$ using a quantum algorithm, e.g, QAOA, and $P_R =1 / 2^n$ is the probability of random guessing with n equal to the number of qubits involved in the problem.

\section{Results}\label{Sec:Results}
\subsection*{Optimal solution distribution}\label{eigenvalues_encoding}
To determine the applicability of our method, we sorted the energy eigenvalues of the cost Hamiltonian for the different combinatorial optimization problems. The parameter of interest is the position of the energy eigenvalue that describes the optimal solution of the original combinatorial optimization problems within the list of all sorted energy eigenvalues of the cost Hamiltonian of their QUBO representation. If the position of the optimal solution is far from the ground energy of the cost Hamiltonian, the unbalanced penalization encoding is not working as expected. Otherwise, if the energy eigenvalue corresponding to the optimal solution is close to the ground state energy given the total number of eigenvalues, the unbalanced encoding is advantageous.

\begin{figure}[!tbh]
\centering
\includegraphics[width=8.5cm]{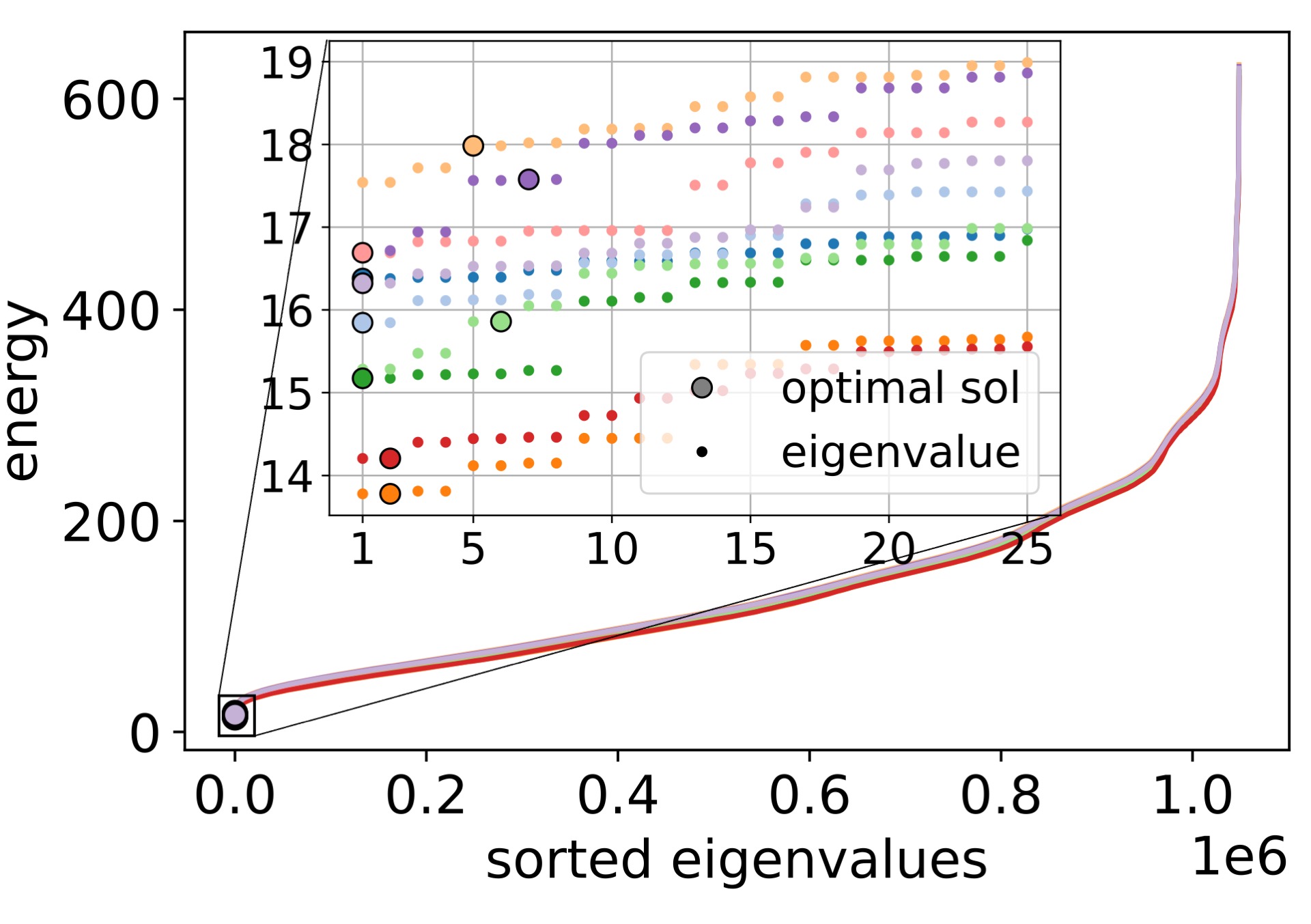}
\caption{\label{eigenenergies_tsp} Eigenvalues distribution for the TSP with 5 cities (20 qubits) for 10 randomly generated problems using the unbalanced penalization method. The inset shows the lowest 25 energy eigenvalues. The big circles are the optimal solutions for the random problem and the small circles are the different eigenvalues of the cost Hamiltonian of its QUBO representation. Note that each eigenvalue is degenerate with multiplicity two, because the problem is symmetric, e.g., one clockwise solution and another anti-clockwise. }
\end{figure}

Figure \ref{eigenenergies_tsp} shows the energy eigenvalues of 10 random cases for the TSP with 5 nodes using the unbalanced penalization encoding approach. The eigenvalues are sorted in terms of their energy from minimum to maximum with the x-axis representing the eigenvalue position (number) of eigenvalues, and the y-axis its energy. In the inset, showing the lowest 25 energy eigenvalues, the big circles represent where the optimal result is positioned compared to all the other eigenvalues (small circles) of the cost Hamiltonian. Here, the worst eigenvalue position is 7 out of $2^{20}$ = 1048576 eigenvalues. This means that in the worst scenario, the optimal solution is among the lowest 0.00057\% of all eigenvalues.

\begin{figure}[!tbh]
\centering
\includegraphics[width=8.5cm]{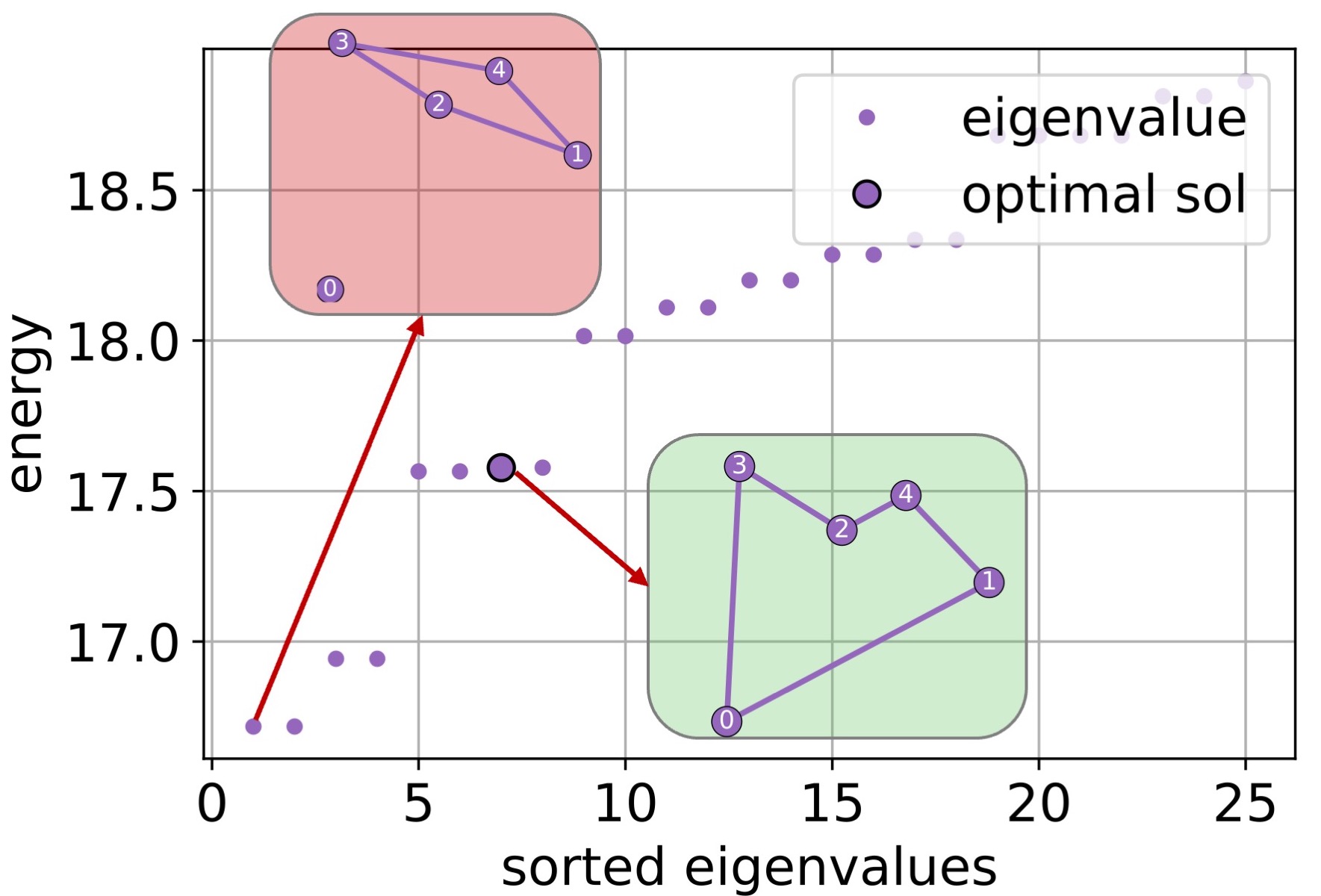}
\caption{\label{suboptimal_tsp} The first 25 sorted eigenvalues of the Hamiltonian $H_c$ corresponding to the worst case shown in Fig.~\ref{eigenenergies_tsp}. The big circle corresponds to the optimal solution of this problem and the small circles are the other eigenvalues. The solution shown in the red inset is the ground state, which is an invalid solution of the TSP. The solution in the green inset is the optimal solution.}
\end{figure}

From Fig. \ref{eigenenergies_tsp} it is seen that the unbalanced penalization approach does not ensure that the optimal solution is the lowest eigenvalue of the cost Hamiltonian, but in all the cases analyzed, it is very close to it. For example, in Fig.~\ref{suboptimal_tsp}, this feature is illustrated for one of the random problems of Fig.~\ref{eigenenergies_tsp}. Here, there are a couple of eigenvalues that do not fulfill all the restrictions of the problem but have lower energy than the optimal solution. In the end, the unbalanced penalization encoding is a trade-off between adding slack variables and ensuring the ground state of the cost Hamiltonian corresponds to the optimal solution of the combinatorial optimization problem, or reducing the number of variables but encoding the optimal solution in the vicinity of the ground state. In many cases, there is no preference for encoding the optimal solution in the lowest eigenvalue of the cost Hamiltonian. For instance, QAOA brings the expectation value of a cost Hamiltonian to a region of overall minimum energy, so we can expect that the optimal and sub-optimal solutions occur with reasonable probability in regions of minimal energy. 

\begin{figure}[!tbh]
\centering
\includegraphics[width=8.5cm]{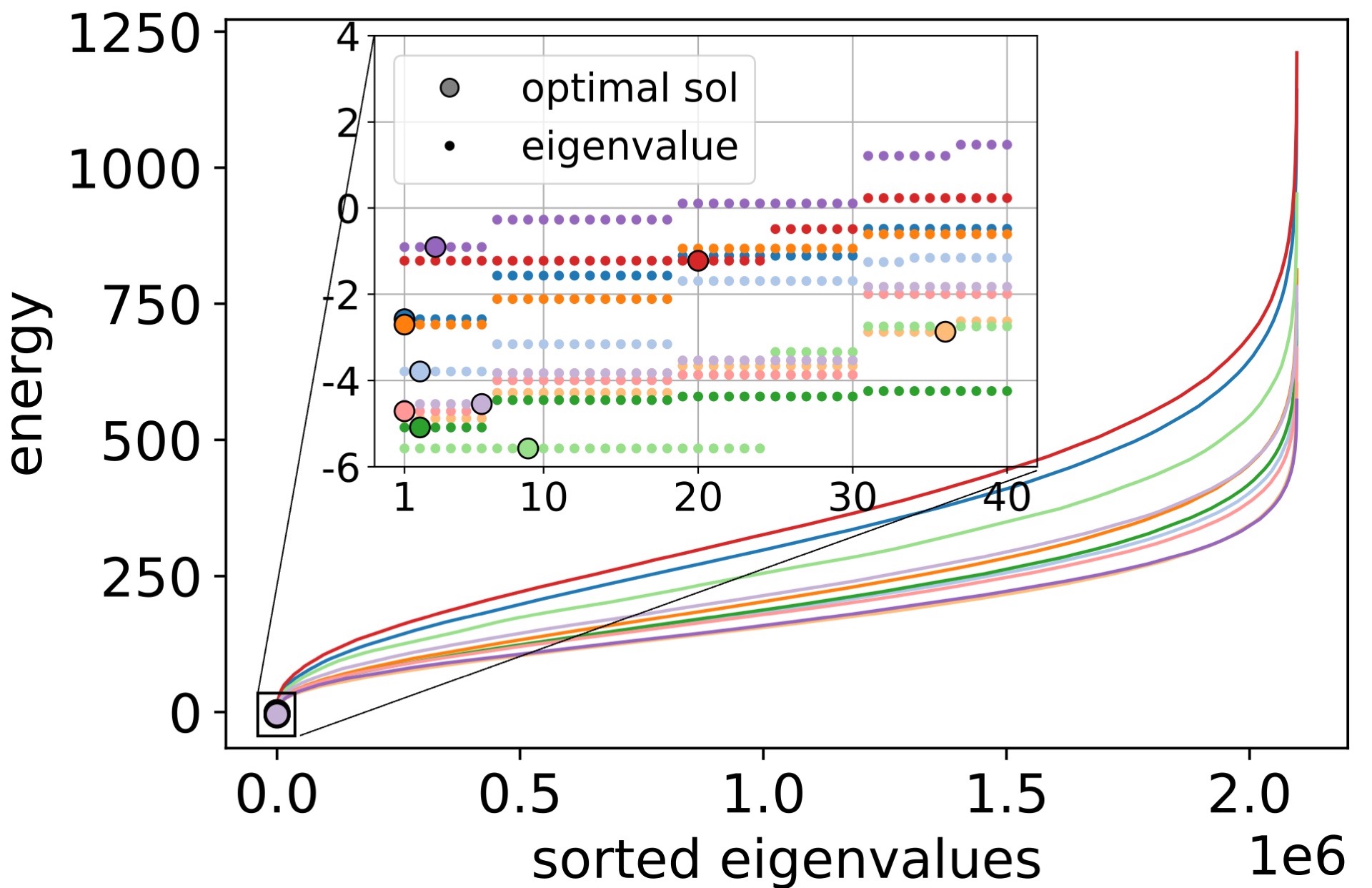}
\caption{\label{eigenenergies_bpp} Eigenvalue distribution for the BPP with 5 items (21 qubits) for 10 randomly generated problems using the unbalanced inequality-constrained penalization method. The inset shows the 40 lowest energy eigenvalues. The big circles are the optimal solutions for the random problem and the small circles are the different eigenvalues for the cost Hamiltonian of its QUBO representation. Note that each eigenvalue has a different degeneracy because the problem has some symmetries.}
\end{figure}

Fig.~\ref{eigenenergies_bpp} shows the energy eigenvalues sorted for the cost Hamiltonian of the BPP for 10 random problems with 5 items. In the inset, showing the 40 lowest eigenvalues, the optimal solutions of the cost Hamiltonians are depicted as big circles and the small circles are the other eigenvalues of the cost Hamiltonian. In this case, there is an increasing number of degeneracies explained by the symmetries of the problem (exchange items of one bin with the others). To keep the same number of qubits (21) for all the problems, we select 10 random cases that require more than 3 bins to store the items. Here, the worst case is in position 36 out of $2^{21}$=2097152 eigenvalues which means it is within $0.00124\%$, of all eigenvalues.

\begin{figure}[!tbh]
\centering
\includegraphics[width=8.5cm]{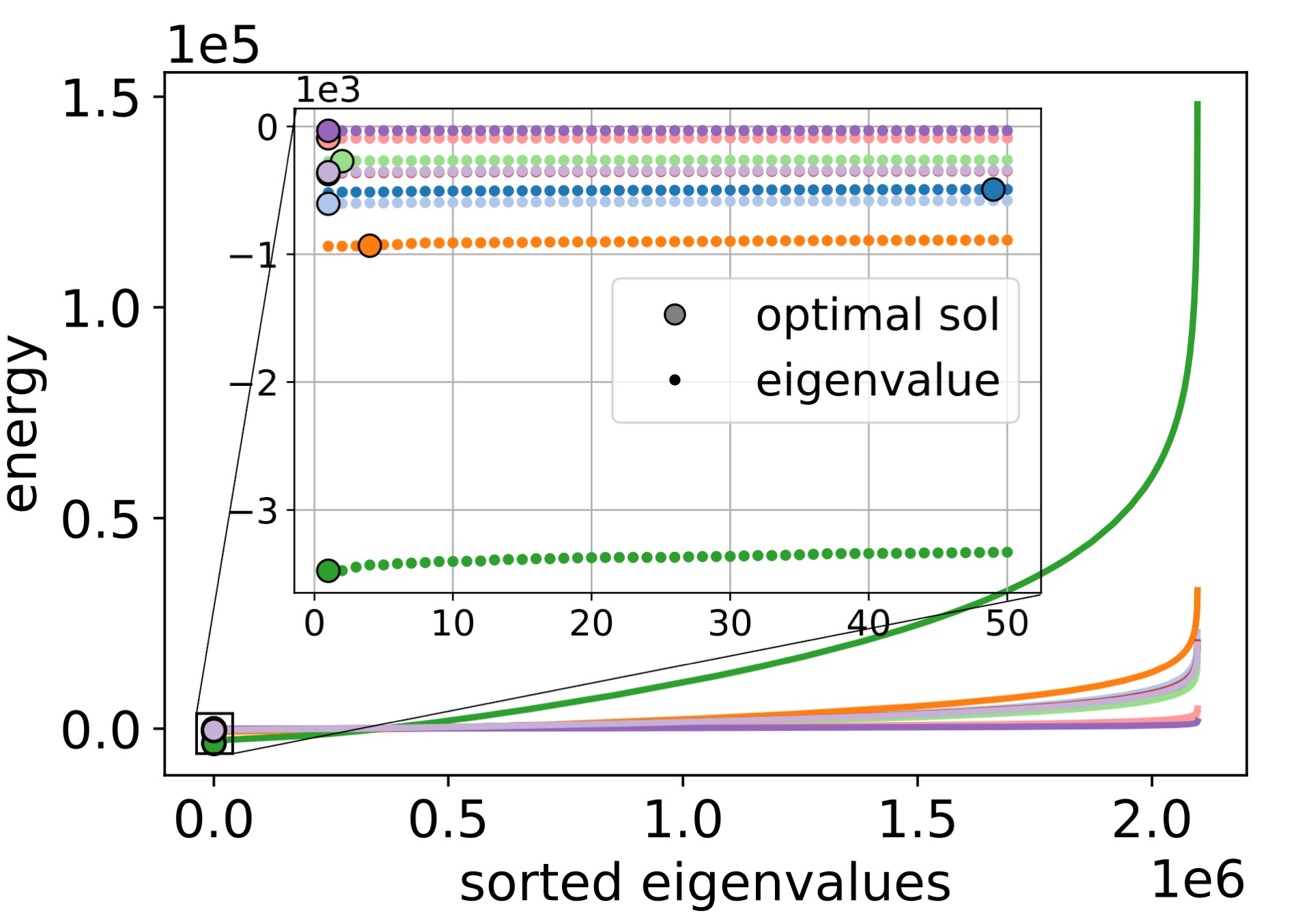}
\caption{\label{eigenenergies_kp} Eigenvalue distribution for the KP with 21 items (21 qubits) for 10 randomly generated problems using the unbalanced penalization method. The inset shows the 50 lowest eigenvalues. The big circles give the position of the optimal solutions for the random problems and the small circles depict the position of the different eigenvalues of the cost Hamiltonian. In this case, there are no degeneracies but the energies are close to each other.}
\end{figure}

Finally, Fig.~\ref{eigenenergies_kp} shows the eigenvalue distribution for 10 random cases of the KP with 21 items (21 qubits). In this case, there are no degeneracies, but the possible solutions are close to each other. The inset shows the first 50 eigenvalues. In this case, for one of the random cases, the optimal solution is located at position 49 out of $2^{21}$=2097152 eigenvalues, which means it is within $0.0023\%$ of the first eigenvalues. In this case, the parameters $\lambda_{0,1,2}$ were optimized for a random case with 13 items, which suggests the great generalization capabilities of the current method.

\subsection*{Quantum Annealing}\label{QA}

\begin{figure}[!tbh]
\centering
\includegraphics[width=8.5cm]{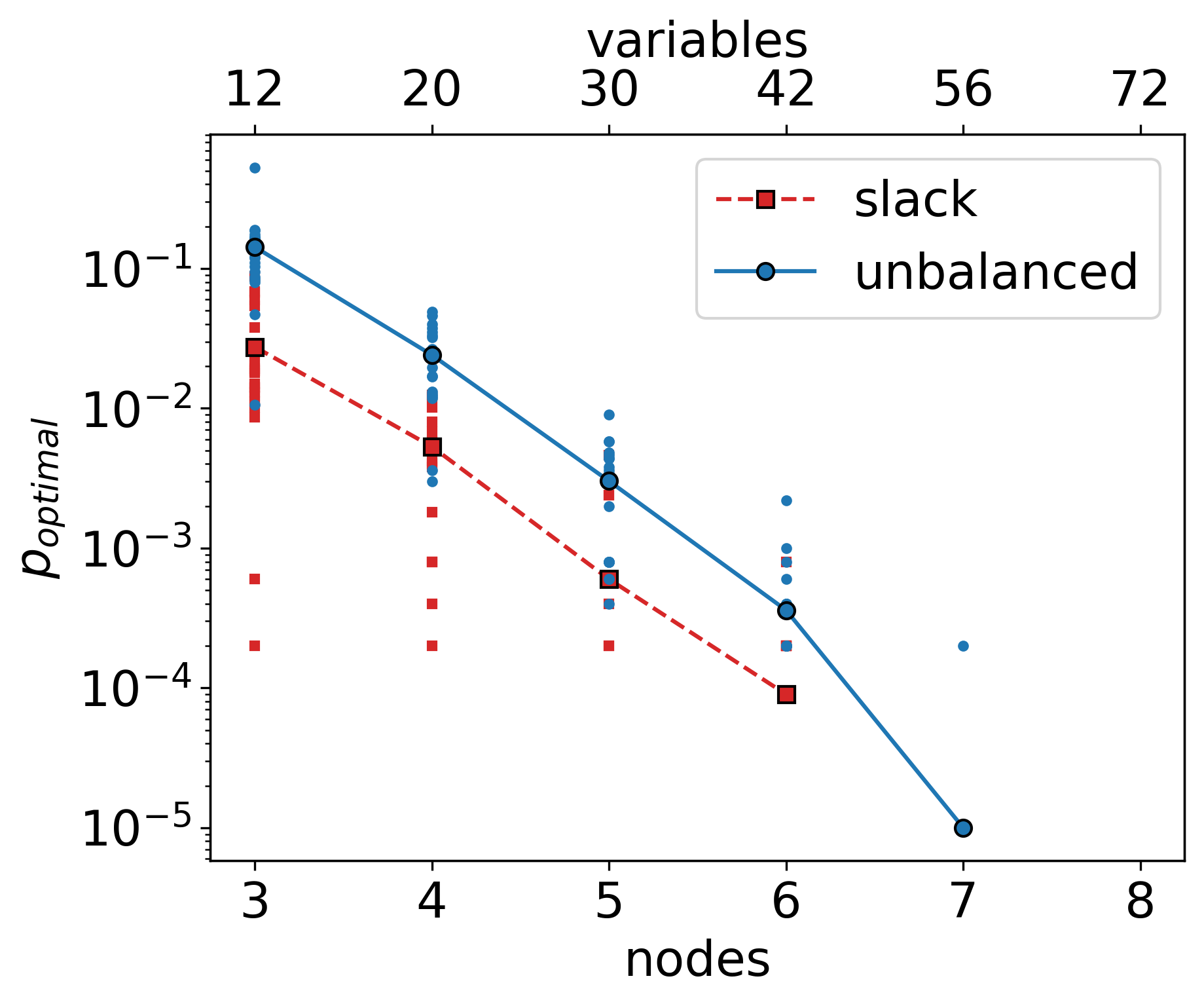}
\caption{\label{QPU-opt} D-Wave Advantage success rate of finding optimal solutions for different cases of the BPP ranging from 3 to 8 nodes (items). The 20 small circles (squares) for each problem size consist of 5000 samples for the unbalanced (slack) method.}
\end{figure}

Figure \ref{QPU-opt} illustrates the results obtained using the D-Wave Advantage. The upper x-axis in this figure, as well as in subsequent plots, represents the logical variables required to represent each problem size, as per the formulation provided in Section \ref{BPP}. Across all cases, the unbalanced penalization approach consistently generates more optimal solutions compared to the slack variables approach. Although the optimal solution cannot be guaranteed to be the ground state of the Hamiltonian, more optimal solutions are found for the unbalanced penalization method.

For the scenario involving 8 nodes, neither of the two methods were able to find valid solutions. However, it is important to acknowledge that the number of statistical samples used in these experiments might not have been sufficient to draw conclusions about the limit beyond which no further solutions are found on the quantum annealer.

\begin{figure}[!tbh]
\centering
\includegraphics[width=8.5cm]{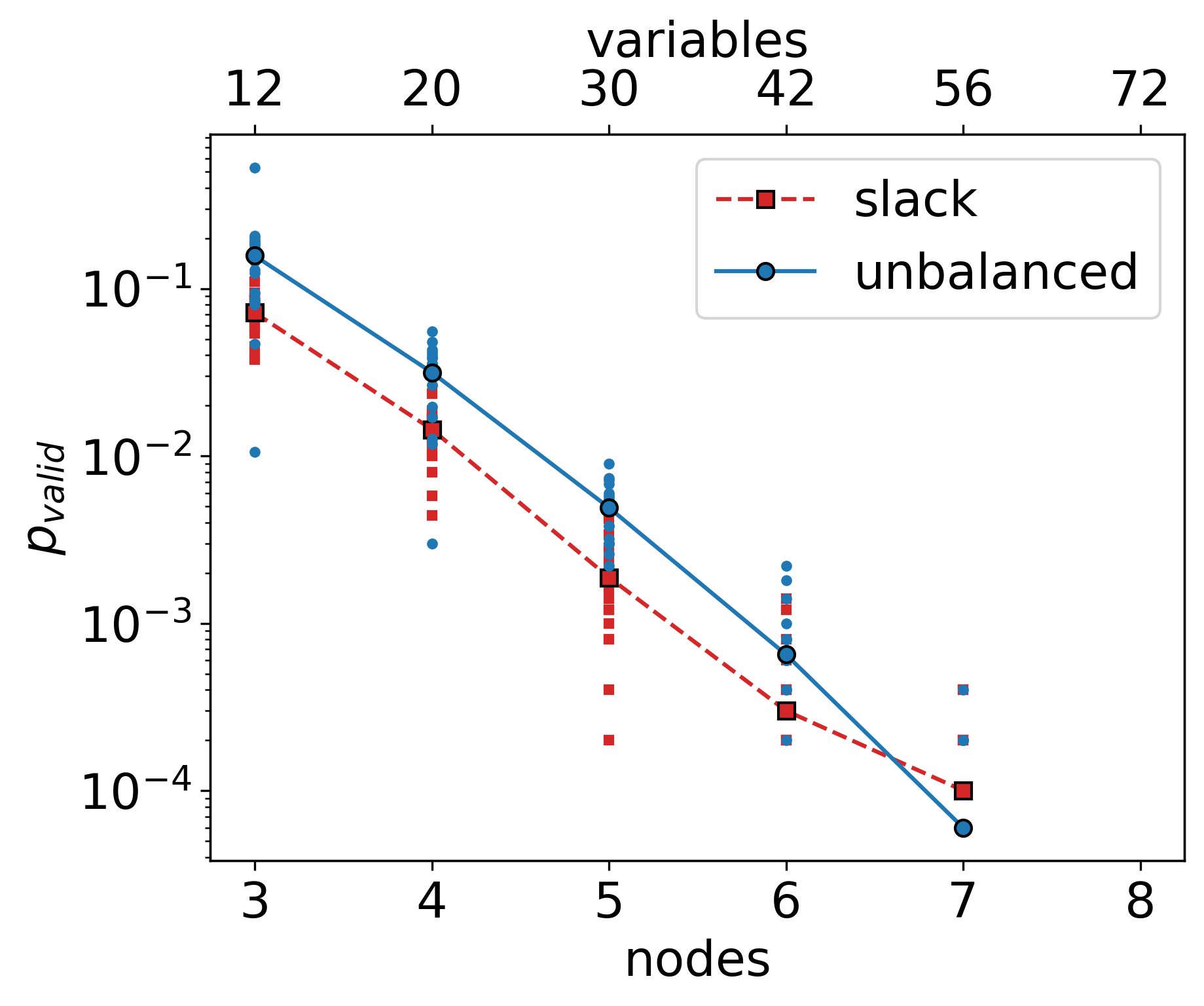}
\caption{\label{QPU-valid} D-Wave Advantage success rate of finding valid solutions for different cases of the BPP ranging from 3 to 8 nodes (items).}
\end{figure}

Fig. \ref{QPU-valid} presents the results of the same experiments depicted in Fig. \ref{QPU-opt}, focusing on the number of valid solutions obtained. Consistent with the observations in Fig. \ref{QPU-opt}, the employment of the unbalanced penalization method enhances the likelihood of discovering valid solutions compared to the slack approach. However, it is important to note that in the case of 7 nodes, where the slack approach appears to exhibit superior performance, more samples of valid solutions (than could be produced by the machine) would be needed to draw a solid conclusion.

In order to investigate the viability of finding valid solutions for larger problem sizes, we employed the D-Wave Hybrid solver. We conducted experiments using problem sizes ranging from 5 to 31 nodes (items), with intervals of 2 and executed 20 random problem instances for each case. Each problem instance was solved using a single run on the D-Wave Hybrid solver, with a maximum time limit of 3 seconds.

\begin{figure}[!tbh]
\centering
\includegraphics[width=8.5cm]{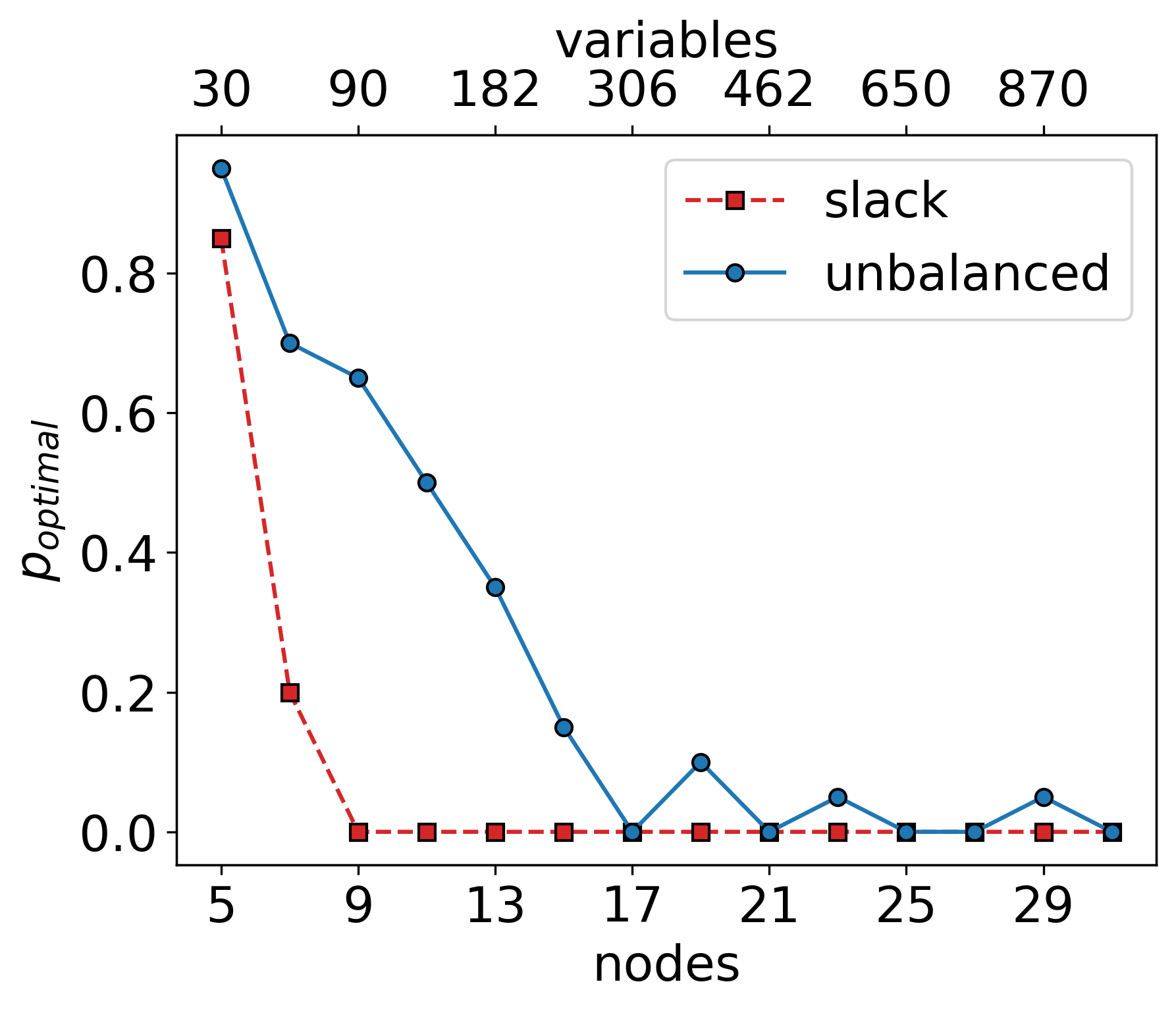}
\caption{\label{HS-opt} D-Wave Hybrid solver success rate of finding optimal solutions for different cases of the BPP ranging from 5 to 31 nodes (items). Each point consists of a single execution of 20 random problems with a maximum time for the hybrid solver of 3 seconds.}
\end{figure}

Figure \ref{HS-opt} depicts the probability of finding optimal solutions across the 20 random problems for each node case. Similar to the observations with the D-Wave Advantage, the unbalanced penalization method consistently outperforms the slack approach. It successfully discovers optimal solutions for nearly all cases, with the exception of 19, 23, 27, and 31 nodes. Conversely, the slack approach fails to find optimal solutions after 7 nodes.

\begin{figure}[!tbh]
\centering
\includegraphics[width=8.5cm]{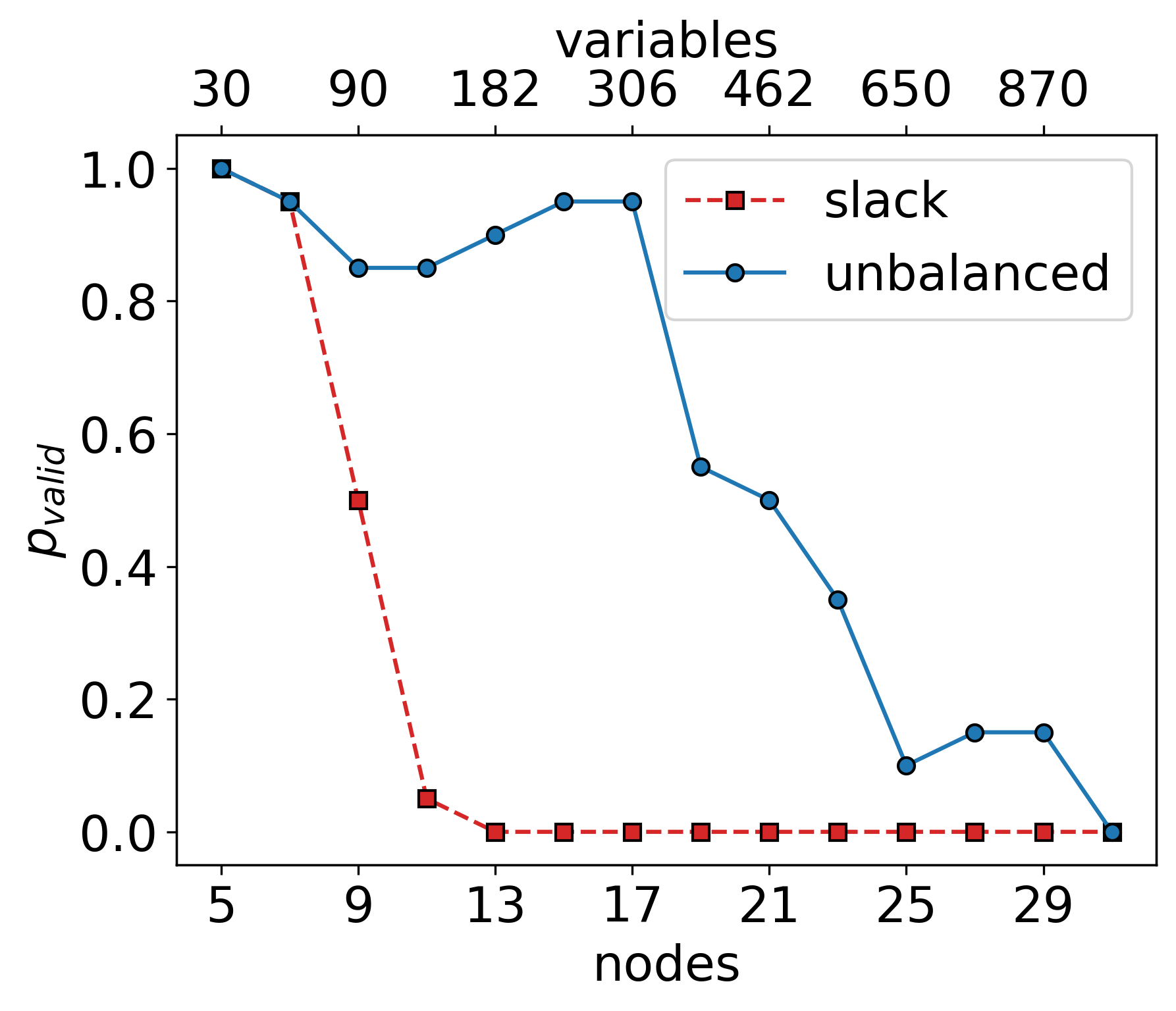}
\caption{\label{HS-valid}  D-Wave Hybrid solver success rate of finding valid solutions for different cases of the BPP ranging from 5 to 31 nodes (items). Each circle (square) represents the mean value of 20 random problems with a maximum time for the hybrid solver of 3 seconds for the unbalanced (slack) method.}
\end{figure}

Figure \ref{HS-valid} illustrates the probability of finding valid solutions using both the unbalanced penalization method and the slack variables approach on the D-Wave Advantage. Consistent with the findings in Figure \ref{HS-opt}, the unbalanced penalization method demonstrates superior performance compared to the slack variables approach.

Up to 17 nodes, the unbalanced penalization method consistently discovers valid solutions, whereas the slack variables approach only manages to find valid solutions up to 11 nodes. Beyond 17 nodes, the probability of finding valid solutions gradually declines for the unbalanced penalization method, with no solutions being found for the 31-node case. This behavior can be attributed to the fact that the maximum time limit remains constant throughout the experiments.

Even though increasing the time will impact positively on improving the performance of both methods, it is important to reiterate that the primary objective of this section is not to provide high-quality solutions but rather to evaluate the performance of both methods under similar conditions.

\subsection*{QAOA landscape}\label{QAOAlandscape}
In this section, the results of the cost Hamiltonian energy landscape using the QAOA algorithm with one layer are presented for the TSP, the BPP, and the KP. The slack variables method and the unbalanced penalization method are used for the cost Hamiltonian encoding. For visualization purposes, the energy landscape for all cases is limited to the region of $\gamma \in (- \pi/\max\{q_{ij}\}$, 0) where $q_{ij}$ is given in Eq.~(\ref{IsingH}) and $\beta \in (-\pi/2, 0)$.

\begin{figure}[!tbh]
\centering
\includegraphics[width=8.5cm]{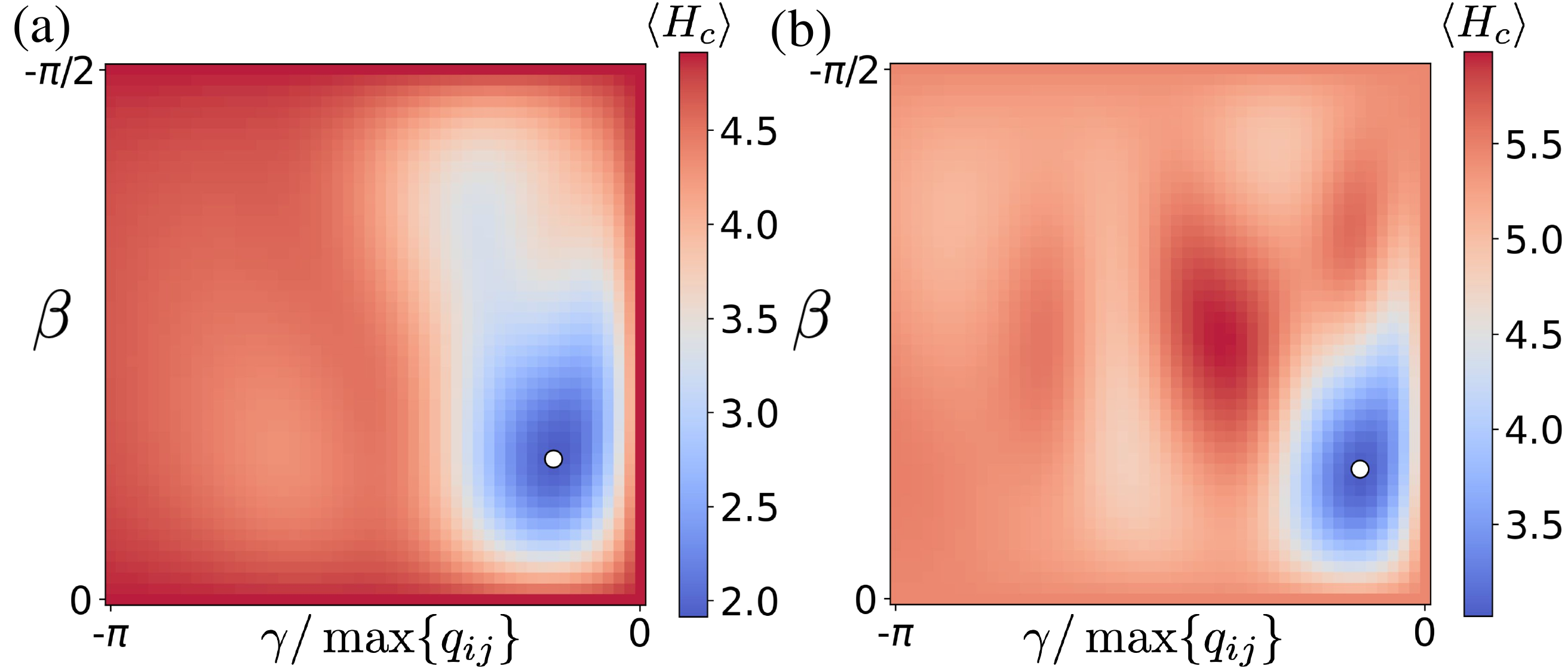}
\caption{\label{unbalanced-TSP} QAOA energy for $p=1$ for the cost Hamiltonian $\langle H_c \rangle$ of the TSP with 4 cities with (a) slack variables encoding and (b) unbalanced penalization encoding. The minimum energy is indicated with a white dot. The probability that this minimum corresponds to the optimal solution of the TSP problem is P(*x) = 0.02\% and P(*x)=0.227\% for encoding (a) and (b), respectively.}
\end{figure}

Fig.~\ref{unbalanced-TSP} shows the energy landscape of the QAOA algorithm for a TSP with 4 cities. This is equivalent to a 17 and 12 qubits problem for the slack variables and the unbalanced penalization encodings, respectively. The probabilities for finding the optimal solutions of the TSP are 0.02\% and 0.227\% for the slack variables and the unbalanced encodings, respectively. This shows a clear advantage of the unbalanced encoding method that increases the probability more than 10 times compared to the probability obtained with the slack variables encoding. This characteristic is mainly attributed to the difference in the number of qubits needed to represent the problem and thus the size of the set of possible solutions. For the slack variables encoding, there are 131072 ($2^{17}$) possible solutions while for the unbalanced penalization encoding, there are 4096 ($2^{12}$)  possible solutions.

\begin{figure}[!tbh]
\centering
\includegraphics[width=9cm]{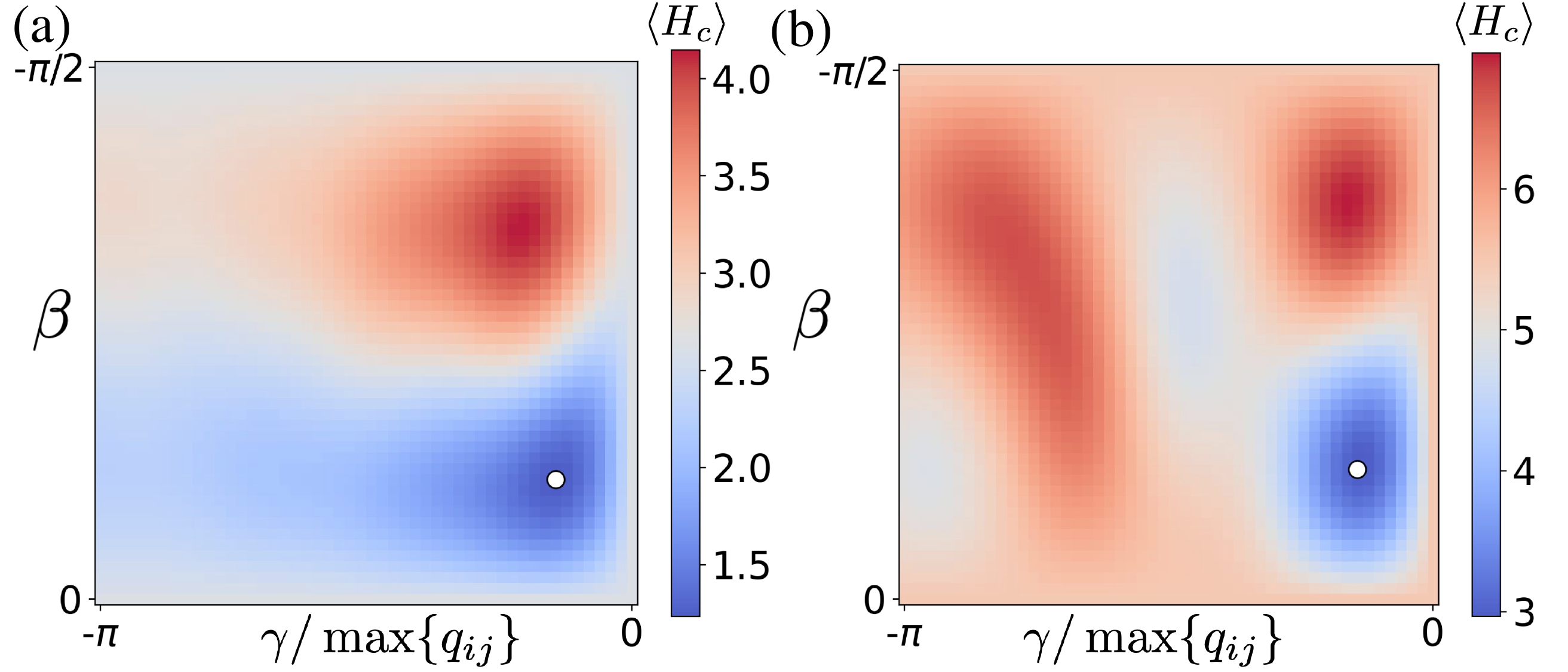}
\caption{\label{unbalanced-BPP} Same as the Fig.~\ref{unbalanced-TSP} for the BPP with 3 items. The probability that this minimum corresponds to the optimal solution of the BPP problem is P(*x) = 0.001\% and P(*x)=10.661\% for encoding (a) and (b), respectively.}
\end{figure}

Fig.~\ref{unbalanced-BPP} shows the energy landscape of the QAOA algorithm for a BPP with 3 items, which is equivalent to a 6 and 19 qubits problem for the unbalanced and slack encodings, respectively. The probability of finding the optimal solution of the BPP in the local minima of QAOA is 0.001\% for the slack variables encoding and 10.66\% for the unbalanced penalization method. Here, the optimal solution is 10000 times more likely to be found using the unbalanced penalization approach than using the slack variables approach.

\begin{figure}[!tbh]
\centering
\includegraphics[width=9 cm]{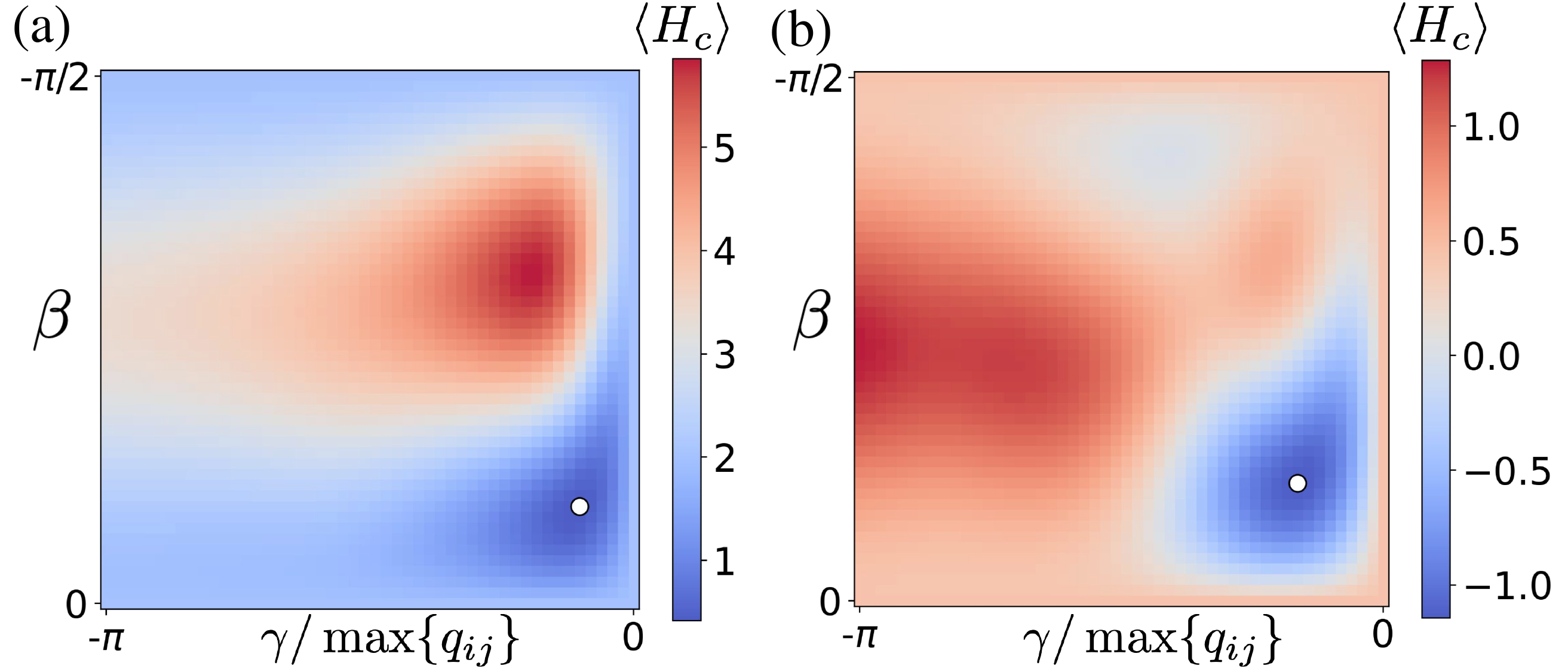}
\caption{\label{unbalanced-knapsack}Same as the Fig.~\ref{unbalanced-TSP} for the KP with 10 items. The probability that this minimum corresponds to the optimal solution of the KP problem is P(*x) = 0.001\% and P(*x)=0.346\% for encoding (a) and (b), respectively.}
\end{figure}

Finally, Fig.~\ref{unbalanced-knapsack} shows the energy landscape of the QAOA algorithm for a KP with 10 items, which is equivalent to an 18 and 10 qubits problem for the slack variables and the unbalanced penalization encoding, respectively. Here, the probability of finding the optimal solution of the KP is 0.001\% and 0.346\% for the slack variables and the unbalanced penalization encoding. This means finding the solution with the unbalanced penalization method is 346 times more likely than with the slack variables encoding.

\subsection*{The coefficient of performance}
\begin{figure*}[!tbh]
\centering
\includegraphics[width=18cm]{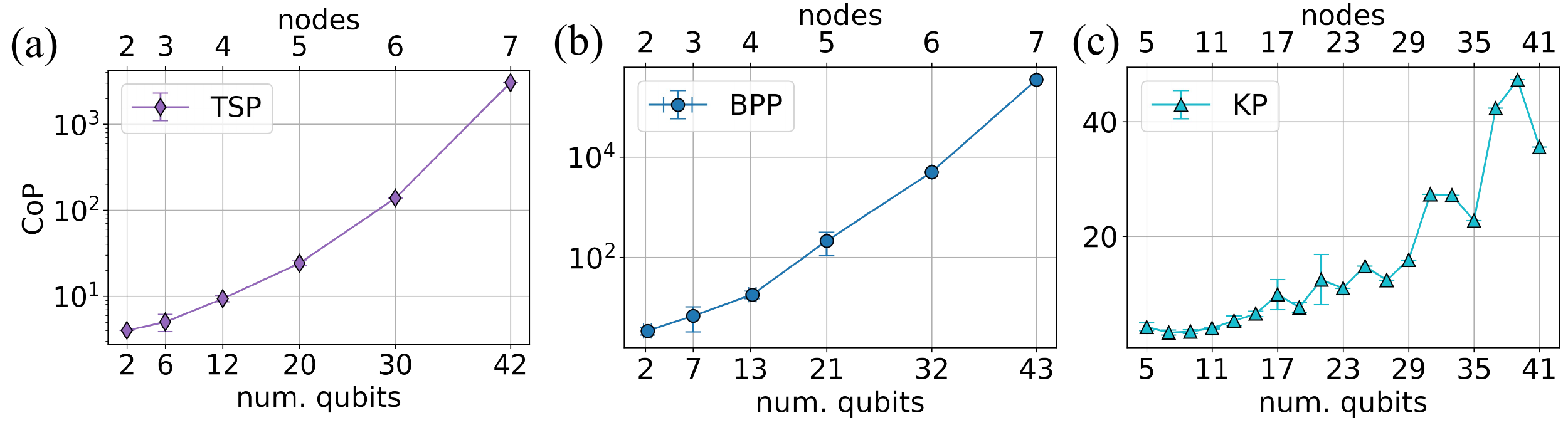}
\caption{\label{CoP} Coefficient of performance for (a) the TSP, (b) the BPP, and (c) the KP for different problem sizes. The error bars represent the standard deviation over 5 different random cases for problems with less than 22 qubits. For problems with more than 21 qubits the data points represent 1 random case.}
\end{figure*}
Figure \ref{CoP} shows the CoP for different problem sizes for the TSP, BPP, and KP using the unbalanced penalization method. All the cases show the probability of finding the optimal solution for each problem using QAOA with 1 layer in the minimum point of the energy landscape region $\beta\in(-\pi/2,0)$ and $\gamma\in(-\pi/\max\{q_{ij}\},0)$. At this point, we do not use a classical optimizer to find the optimal beta and gamma, but instead, we divide the beta and gamma region in a 50x50 grid landscape and take the probability of the optimal solution at the minimum energy. It is expected that using an optimization method for this step improves the CoP while keeping the same tendency. Fig.~\ref{CoP} (a) shows the CoP for the TSP problem with 2, 3, 4, 5, 6, and 7 cities (2, 6, 12, 20, 30, and 42 qubits). For the cases with 30 and 42 qubits, we use JUQCS \cite{DeRaedt2007MassivelyParallel,DeRaedt2018MassivelyParallel} in its GPU-accelerated version \cite{Willsch2021JUQCSGQAOA} to obtain the results. Note that the same TSP cases using the slack variables approach will require 2, 7, 17, 36, 73, 148, and 304 qubits. We thus explore problem sizes (5, 6, and 7 cities) with the unbalanced penalization approach that are unfeasible using the slack approach on quantum computer simulators. Interestingly, the CoP increases exponentially with the problem size, which means that the probability of finding the optimal solution exponentially increases when we compare it with random guessing in the set of possible configurations (cf.~\cite{Willsch2022HybridQuantumClassicalSimulations}).

Fig.~\ref{CoP} (b) shows the CoP for the BPP for 2, 3, 4, 5, 6, and 7 items (2, 6, 13, 21, 32, and 43 qubits). For this case, the CoP is improving even more rapidly compared to TSP and also here for the BPP with 6 and 7 items, the solutions using slack variables (61 and 78 qubits) are beyond the range of what can be simulated with quantum computer simulators. The largest simulations up to 43 qubits were performed on JUWELS Booster \cite{JUWELSBooster} as they required more than $1/8\,\mathrm{PiB}$ of distributed memory and took more than one million core hours. Finally, Fig.~\ref{CoP} (c) shows the CoP for the KP with 5 to 41 items (5 to 41 qubits) with increments of two (same number of qubits). In this case, the CoP increases linearly with the number of qubits. This is a poor performance if we compare it with the TSP and BPP. We suspect the improved CoP is related to the number of constraints of the problem.

\section{Conclusions}\label{Sec:Conclusion}

We have presented {\it unbalanced penalization}, a new method to encode inequality constraints of combinatorial optimization problems into QUBO penalizations. The method does not require extra slack variables to encode the inequality constraints. This is extremely beneficial because there is no increase in the number of variables or qubits needed to encode the problems, especially for the TSP and the BPP. 

The method is suitable for VQA on gate-based quantum computers and QA on quantum annealers, which both require expressing the combinatorial optimization problems in terms of {QUBOs}. We have tested the method using QA for the BPP and QAOA with 1 layer for the TSP, BPP, and KP. We have shown that in 10 random cases for different problem sizes the optimal solution of the combinatorial optimization problem is located in the vicinity of the ground state energy of the cost Hamiltonian. The method is highly generalizable in the sense that the tuned parameters $\lambda_{0,1,2}$ for small problems generalize well to larger problems.
For QA, we have demonstrated that the unbalanced penalization method outperforms the slack variables approach, finding better solutions in terms of quality and quantity. Additionally, the probabilities to find the optimal solutions among the local minima of the unbalanced penalization (Figs.~\ref{unbalanced-TSP} to \ref{unbalanced-knapsack}) are much better than using slack variables. Qualitatively, this can be understood because the addition of slack variables increases the search space and therefore reduces the probability of finding optimal and nearly-optimal solutions. 

The proposed method was evaluated on both the D-Wave Advantage and D-Wave Hybrid solvers. Our findings demonstrate that the method utilizing unbalanced penalization consistently outperforms the slack variables approach in solving the BPP. Specifically, the unbalanced penalization method successfully finds solutions for problem instances with up to 31 items using the Hybrid solver, while the slack variables approach achieves solutions for only up to 11 items. We remark that additional evaluations of this method on D-Wave systems for the TSP support the same conclusions \cite{Montanez-Barrera2023}.

We have tested the unbalanced penalization method close to the limits of what is possible to be simulated on supercomputers using JUQCS. For the case of the TSP, we went up to 42 qubits, for the BPP up to 43 qubits, and for the KP up to 41 qubits. Indeed, we are exploring regions unfeasible using the slack variables method. For example, for the TSP with 6 and 7 cities using the slack variables method it requires 73 and 148 qubits, and for the BPP with 6 and 7 items it requires 61 and 78 qubits. In terms of probability, the unbalanced penalization method shows large advantages for the small problems we made a comparison for. For the TSP with 4 cities, the probability of finding the optimal solution at the local minimum energy of the QAOA with $p=1$ using the unbalanced penalization method is more than 10 times larger than the one using the slack variables approach. For the BPP with 3 items, the probability of finding the optimal solution with the unbalanced penalization method is more than 10000 times larger, and for the KP with 10 items, it is almost 346 higher.

Finally, we have presented a new metric, the CoP, which compares the probability of obtaining a specific solution using a quantum algorithm against random guessing over the set of possible solutions. The CoP is useful to characterize how an algorithm performs with increasing problem size or to compare how different combinatorial optimization problems perform for a specific algorithm. We have shown the CoP for the TSP, BPP, and KP using QAOA with $p=1$ in the region $\gamma \in (- \pi/\max\{q_{ij}\}$, 0) and $\beta \in (- \pi/2$, 0). Interestingly, for all of them, the CoP increases with the problem size. For the KP the increase is linear while for the TSP and BPP, the increase is exponential. Further research is worthwhile to see if this provides an implicit advantage of using QAOA to solve combinatorial optimization problems, what the characteristics of the TSP and the BPP are that make the CoP increase exponentially, and to see how increasing the number of QAOA layers improves the CoP.

\begin{acknowledgments}
\vspace{-10pt}
We would like to thank Leonardo Disilvestro and Vishal Sharma from Entropica Labs for their helpful comments and suggestions about the results of the present work.
J. A. Monta\~nez-Barrera acknowledges support by the German Federal Ministry of Education and Research (BMBF), funding program Quantum technologies - from basic research to market, project QSolid (Grant No. 13N16149).
D.~Willsch acknowledges support from the project J\"ulich UNified Infrastructure for Quantum computing (JUNIQ) that has received funding from the German Federal Ministry of Education and Research (BMBF) and the Ministry of Culture and Science of the State of North Rhine-Westphalia.
The authors gratefully acknowledge the Gauss Centre for Supercomputing e.V. (www.gauss-centre.eu) for funding this project by providing computing time on the GCS Supercomputer JUWELS at Jülich Supercomputing Centre (JSC).

\end{acknowledgments}
\clearpage  

\bibliography{References}

\begin{thebibliography}{54}%
\makeatletter
\providecommand \@ifxundefined [1]{%
 \@ifx{#1\undefined}
}%
\providecommand \@ifnum [1]{%
 \ifnum #1\expandafter \@firstoftwo
 \else \expandafter \@secondoftwo
 \fi
}%
\providecommand \@ifx [1]{%
 \ifx #1\expandafter \@firstoftwo
 \else \expandafter \@secondoftwo
 \fi
}%
\providecommand \natexlab [1]{#1}%
\providecommand \enquote  [1]{``#1''}%
\providecommand \bibnamefont  [1]{#1}%
\providecommand \bibfnamefont [1]{#1}%
\providecommand \citenamefont [1]{#1}%
\providecommand \href@noop [0]{\@secondoftwo}%
\providecommand \href [0]{\begingroup \@sanitize@url \@href}%
\providecommand \@href[1]{\@@startlink{#1}\@@href}%
\providecommand \@@href[1]{\endgroup#1\@@endlink}%
\providecommand \@sanitize@url [0]{\catcode `\\12\catcode `\$12\catcode
  `\&12\catcode `\#12\catcode `\^12\catcode `\_12\catcode `\%12\relax}%
\providecommand \@@startlink[1]{}%
\providecommand \@@endlink[0]{}%
\providecommand \url  [0]{\begingroup\@sanitize@url \@url }%
\providecommand \@url [1]{\endgroup\@href {#1}{\urlprefix }}%
\providecommand \urlprefix  [0]{URL }%
\providecommand \Eprint [0]{\href }%
\providecommand \doibase [0]{https://doi.org/}%
\providecommand \selectlanguage [0]{\@gobble}%
\providecommand \bibinfo  [0]{\@secondoftwo}%
\providecommand \bibfield  [0]{\@secondoftwo}%
\providecommand \translation [1]{[#1]}%
\providecommand \BibitemOpen [0]{}%
\providecommand \bibitemStop [0]{}%
\providecommand \bibitemNoStop [0]{.\EOS\space}%
\providecommand \EOS [0]{\spacefactor3000\relax}%
\providecommand \BibitemShut  [1]{\csname bibitem#1\endcsname}%
\let\auto@bib@innerbib\@empty
\bibitem [{\citenamefont {Lucas}(2014)}]{Lucas2014}%
  \BibitemOpen
  \bibfield  {author} {\bibinfo {author} {\bibfnamefont {A.}~\bibnamefont
  {Lucas}},\ }\bibfield  {title} {\bibinfo {title} {{Ising formulations of many
  NP problems}},\ }\href {https://doi.org/10.3389/fphy.2014.00005} {\bibfield
  {journal} {\bibinfo  {journal} {Frontiers in Physics}\ }\textbf {\bibinfo
  {volume} {2}},\ \bibinfo {pages} {1} (\bibinfo {year} {2014})},\ \Eprint
  {https://arxiv.org/abs/1302.5843} {arXiv:1302.5843} \BibitemShut {NoStop}%
\bibitem [{\citenamefont {Kochenberger}\ \emph {et~al.}(2014)\citenamefont
  {Kochenberger}, \citenamefont {Hao}, \citenamefont {Glover}, \citenamefont
  {Lewis}, \citenamefont {L{\"{u}}}, \citenamefont {Wang},\ and\ \citenamefont
  {Wang}}]{Kochenberger2014}%
  \BibitemOpen
  \bibfield  {author} {\bibinfo {author} {\bibfnamefont {G.}~\bibnamefont
  {Kochenberger}}, \bibinfo {author} {\bibfnamefont {J.~K.}\ \bibnamefont
  {Hao}}, \bibinfo {author} {\bibfnamefont {F.}~\bibnamefont {Glover}},
  \bibinfo {author} {\bibfnamefont {M.}~\bibnamefont {Lewis}}, \bibinfo
  {author} {\bibfnamefont {Z.}~\bibnamefont {L{\"{u}}}}, \bibinfo {author}
  {\bibfnamefont {H.}~\bibnamefont {Wang}},\ and\ \bibinfo {author}
  {\bibfnamefont {Y.}~\bibnamefont {Wang}},\ }\bibfield  {title} {\bibinfo
  {title} {{The unconstrained binary quadratic programming problem: A
  survey}},\ }\href {https://doi.org/10.1007/s10878-014-9734-0} {\bibfield
  {journal} {\bibinfo  {journal} {Journal of Combinatorial Optimization}\
  }\textbf {\bibinfo {volume} {28}},\ \bibinfo {pages} {58} (\bibinfo {year}
  {2014})}\BibitemShut {NoStop}%
\bibitem [{\citenamefont {Ohzeki}(2020)}]{Ohzeki2020}%
  \BibitemOpen
  \bibfield  {author} {\bibinfo {author} {\bibfnamefont {M.}~\bibnamefont
  {Ohzeki}},\ }\bibfield  {title} {\bibinfo {title} {{Breaking limitation of
  quantum annealer in solving optimization problems under constraints}},\
  }\href {https://doi.org/10.1038/s41598-020-60022-5} {\bibfield  {journal}
  {\bibinfo  {journal} {Scientific Reports}\ }\textbf {\bibinfo {volume}
  {10}},\ \bibinfo {pages} {1} (\bibinfo {year} {2020})},\ \Eprint
  {https://arxiv.org/abs/2002.05298} {arXiv:2002.05298} \BibitemShut {NoStop}%
\bibitem [{\citenamefont {Harrigan}\ \emph {et~al.}(2021)\citenamefont
  {Harrigan}, \citenamefont {Sung}, \citenamefont {Neeley}, \citenamefont
  {Satzinger}, \citenamefont {Arute}, \citenamefont {Arya}, \citenamefont
  {Atalaya}, \citenamefont {Bardin}, \citenamefont {Barends}, \citenamefont
  {Boixo}, \citenamefont {Broughton}, \citenamefont {Buckley}, \citenamefont
  {Buell}, \citenamefont {Burkett}, \citenamefont {Bushnell}, \citenamefont
  {Chen}, \citenamefont {Chen}, \citenamefont {{Ben Chiaro}}, \citenamefont
  {Collins}, \citenamefont {Courtney}, \citenamefont {Demura}, \citenamefont
  {Dunsworth}, \citenamefont {Eppens}, \citenamefont {Fowler}, \citenamefont
  {Foxen}, \citenamefont {Gidney}, \citenamefont {Giustina}, \citenamefont
  {Graff}, \citenamefont {Habegger}, \citenamefont {Ho}, \citenamefont {Hong},
  \citenamefont {Huang}, \citenamefont {Ioffe}, \citenamefont {Isakov},
  \citenamefont {Jeffrey}, \citenamefont {Jiang}, \citenamefont {Jones},
  \citenamefont {Kafri}, \citenamefont {Kechedzhi}, \citenamefont {Kelly},
  \citenamefont {Kim}, \citenamefont {Klimov}, \citenamefont {Korotkov},
  \citenamefont {Kostritsa}, \citenamefont {Landhuis}, \citenamefont {Laptev},
  \citenamefont {Lindmark}, \citenamefont {Leib}, \citenamefont {Martin},
  \citenamefont {Martinis}, \citenamefont {McClean}, \citenamefont {McEwen},
  \citenamefont {Megrant}, \citenamefont {Mi}, \citenamefont {Mohseni},
  \citenamefont {Mruczkiewicz}, \citenamefont {Mutus}, \citenamefont {Naaman},
  \citenamefont {Neill}, \citenamefont {Neukart}, \citenamefont {Niu},
  \citenamefont {O'Brien}, \citenamefont {O'Gorman}, \citenamefont {Ostby},
  \citenamefont {Petukhov}, \citenamefont {Putterman}, \citenamefont
  {Quintana}, \citenamefont {Roushan}, \citenamefont {Rubin}, \citenamefont
  {Sank}, \citenamefont {Skolik}, \citenamefont {Smelyanskiy}, \citenamefont
  {Strain}, \citenamefont {Streif}, \citenamefont {Szalay}, \citenamefont
  {Vainsencher}, \citenamefont {White}, \citenamefont {Yao}, \citenamefont
  {Yeh}, \citenamefont {Zalcman}, \citenamefont {Zhou}, \citenamefont {Neven},
  \citenamefont {Bacon}, \citenamefont {Lucero}, \citenamefont {Farhi},\ and\
  \citenamefont {Babbush}}]{Harrigan2021}%
  \BibitemOpen
  \bibfield  {author} {\bibinfo {author} {\bibfnamefont {M.~P.}\ \bibnamefont
  {Harrigan}}, \bibinfo {author} {\bibfnamefont {K.~J.}\ \bibnamefont {Sung}},
  \bibinfo {author} {\bibfnamefont {M.}~\bibnamefont {Neeley}}, \bibinfo
  {author} {\bibfnamefont {K.~J.}\ \bibnamefont {Satzinger}}, \bibinfo {author}
  {\bibfnamefont {F.}~\bibnamefont {Arute}}, \bibinfo {author} {\bibfnamefont
  {K.}~\bibnamefont {Arya}}, \bibinfo {author} {\bibfnamefont {J.}~\bibnamefont
  {Atalaya}}, \bibinfo {author} {\bibfnamefont {J.~C.}\ \bibnamefont {Bardin}},
  \bibinfo {author} {\bibfnamefont {R.}~\bibnamefont {Barends}}, \bibinfo
  {author} {\bibfnamefont {S.}~\bibnamefont {Boixo}}, \bibinfo {author}
  {\bibfnamefont {M.}~\bibnamefont {Broughton}}, \bibinfo {author}
  {\bibfnamefont {B.~B.}\ \bibnamefont {Buckley}}, \bibinfo {author}
  {\bibfnamefont {D.~A.}\ \bibnamefont {Buell}}, \bibinfo {author}
  {\bibfnamefont {B.}~\bibnamefont {Burkett}}, \bibinfo {author} {\bibfnamefont
  {N.}~\bibnamefont {Bushnell}}, \bibinfo {author} {\bibfnamefont
  {Y.}~\bibnamefont {Chen}}, \bibinfo {author} {\bibfnamefont {Z.}~\bibnamefont
  {Chen}}, \bibinfo {author} {\bibnamefont {{Ben Chiaro}}}, \bibinfo {author}
  {\bibfnamefont {R.}~\bibnamefont {Collins}}, \bibinfo {author} {\bibfnamefont
  {W.}~\bibnamefont {Courtney}}, \bibinfo {author} {\bibfnamefont
  {S.}~\bibnamefont {Demura}}, \bibinfo {author} {\bibfnamefont
  {A.}~\bibnamefont {Dunsworth}}, \bibinfo {author} {\bibfnamefont
  {D.}~\bibnamefont {Eppens}}, \bibinfo {author} {\bibfnamefont
  {A.}~\bibnamefont {Fowler}}, \bibinfo {author} {\bibfnamefont
  {B.}~\bibnamefont {Foxen}}, \bibinfo {author} {\bibfnamefont
  {C.}~\bibnamefont {Gidney}}, \bibinfo {author} {\bibfnamefont
  {M.}~\bibnamefont {Giustina}}, \bibinfo {author} {\bibfnamefont
  {R.}~\bibnamefont {Graff}}, \bibinfo {author} {\bibfnamefont
  {S.}~\bibnamefont {Habegger}}, \bibinfo {author} {\bibfnamefont
  {A.}~\bibnamefont {Ho}}, \bibinfo {author} {\bibfnamefont {S.}~\bibnamefont
  {Hong}}, \bibinfo {author} {\bibfnamefont {T.}~\bibnamefont {Huang}},
  \bibinfo {author} {\bibfnamefont {L.~B.}\ \bibnamefont {Ioffe}}, \bibinfo
  {author} {\bibfnamefont {S.~V.}\ \bibnamefont {Isakov}}, \bibinfo {author}
  {\bibfnamefont {E.}~\bibnamefont {Jeffrey}}, \bibinfo {author} {\bibfnamefont
  {Z.}~\bibnamefont {Jiang}}, \bibinfo {author} {\bibfnamefont
  {C.}~\bibnamefont {Jones}}, \bibinfo {author} {\bibfnamefont
  {D.}~\bibnamefont {Kafri}}, \bibinfo {author} {\bibfnamefont
  {K.}~\bibnamefont {Kechedzhi}}, \bibinfo {author} {\bibfnamefont
  {J.}~\bibnamefont {Kelly}}, \bibinfo {author} {\bibfnamefont
  {S.}~\bibnamefont {Kim}}, \bibinfo {author} {\bibfnamefont {P.~V.}\
  \bibnamefont {Klimov}}, \bibinfo {author} {\bibfnamefont {A.~N.}\
  \bibnamefont {Korotkov}}, \bibinfo {author} {\bibfnamefont {F.}~\bibnamefont
  {Kostritsa}}, \bibinfo {author} {\bibfnamefont {D.}~\bibnamefont {Landhuis}},
  \bibinfo {author} {\bibfnamefont {P.}~\bibnamefont {Laptev}}, \bibinfo
  {author} {\bibfnamefont {M.}~\bibnamefont {Lindmark}}, \bibinfo {author}
  {\bibfnamefont {M.}~\bibnamefont {Leib}}, \bibinfo {author} {\bibfnamefont
  {O.}~\bibnamefont {Martin}}, \bibinfo {author} {\bibfnamefont {J.~M.}\
  \bibnamefont {Martinis}}, \bibinfo {author} {\bibfnamefont {J.~R.}\
  \bibnamefont {McClean}}, \bibinfo {author} {\bibfnamefont {M.}~\bibnamefont
  {McEwen}}, \bibinfo {author} {\bibfnamefont {A.}~\bibnamefont {Megrant}},
  \bibinfo {author} {\bibfnamefont {X.}~\bibnamefont {Mi}}, \bibinfo {author}
  {\bibfnamefont {M.}~\bibnamefont {Mohseni}}, \bibinfo {author} {\bibfnamefont
  {W.}~\bibnamefont {Mruczkiewicz}}, \bibinfo {author} {\bibfnamefont
  {J.}~\bibnamefont {Mutus}}, \bibinfo {author} {\bibfnamefont
  {O.}~\bibnamefont {Naaman}}, \bibinfo {author} {\bibfnamefont
  {C.}~\bibnamefont {Neill}}, \bibinfo {author} {\bibfnamefont
  {F.}~\bibnamefont {Neukart}}, \bibinfo {author} {\bibfnamefont {M.~Y.}\
  \bibnamefont {Niu}}, \bibinfo {author} {\bibfnamefont {T.~E.}\ \bibnamefont
  {O'Brien}}, \bibinfo {author} {\bibfnamefont {B.}~\bibnamefont {O'Gorman}},
  \bibinfo {author} {\bibfnamefont {E.}~\bibnamefont {Ostby}}, \bibinfo
  {author} {\bibfnamefont {A.}~\bibnamefont {Petukhov}}, \bibinfo {author}
  {\bibfnamefont {H.}~\bibnamefont {Putterman}}, \bibinfo {author}
  {\bibfnamefont {C.}~\bibnamefont {Quintana}}, \bibinfo {author}
  {\bibfnamefont {P.}~\bibnamefont {Roushan}}, \bibinfo {author} {\bibfnamefont
  {N.~C.}\ \bibnamefont {Rubin}}, \bibinfo {author} {\bibfnamefont
  {D.}~\bibnamefont {Sank}}, \bibinfo {author} {\bibfnamefont {A.}~\bibnamefont
  {Skolik}}, \bibinfo {author} {\bibfnamefont {V.}~\bibnamefont {Smelyanskiy}},
  \bibinfo {author} {\bibfnamefont {D.}~\bibnamefont {Strain}}, \bibinfo
  {author} {\bibfnamefont {M.}~\bibnamefont {Streif}}, \bibinfo {author}
  {\bibfnamefont {M.}~\bibnamefont {Szalay}}, \bibinfo {author} {\bibfnamefont
  {A.}~\bibnamefont {Vainsencher}}, \bibinfo {author} {\bibfnamefont
  {T.}~\bibnamefont {White}}, \bibinfo {author} {\bibfnamefont {Z.~J.}\
  \bibnamefont {Yao}}, \bibinfo {author} {\bibfnamefont {P.}~\bibnamefont
  {Yeh}}, \bibinfo {author} {\bibfnamefont {A.}~\bibnamefont {Zalcman}},
  \bibinfo {author} {\bibfnamefont {L.}~\bibnamefont {Zhou}}, \bibinfo {author}
  {\bibfnamefont {H.}~\bibnamefont {Neven}}, \bibinfo {author} {\bibfnamefont
  {D.}~\bibnamefont {Bacon}}, \bibinfo {author} {\bibfnamefont
  {E.}~\bibnamefont {Lucero}}, \bibinfo {author} {\bibfnamefont
  {E.}~\bibnamefont {Farhi}},\ and\ \bibinfo {author} {\bibfnamefont
  {R.}~\bibnamefont {Babbush}},\ }\bibfield  {title} {\bibinfo {title}
  {{Quantum approximate optimization of non-planar graph problems on a planar
  superconducting processor}},\ }\href
  {https://doi.org/10.1038/s41567-020-01105-y} {\bibfield  {journal} {\bibinfo
  {journal} {Nature Physics}\ }\textbf {\bibinfo {volume} {17}},\ \bibinfo
  {pages} {332} (\bibinfo {year} {2021})},\ \Eprint
  {https://arxiv.org/abs/2004.04197} {2004.04197} \BibitemShut {NoStop}%
\bibitem [{\citenamefont {Niroula}\ \emph {et~al.}(2022)\citenamefont
  {Niroula}, \citenamefont {Shaydulin}, \citenamefont {Yalovetzky},
  \citenamefont {Minssen}, \citenamefont {Herman}, \citenamefont {Hu},\ and\
  \citenamefont {Pistoia}}]{Niroula2022}%
  \BibitemOpen
  \bibfield  {author} {\bibinfo {author} {\bibfnamefont {P.}~\bibnamefont
  {Niroula}}, \bibinfo {author} {\bibfnamefont {R.}~\bibnamefont {Shaydulin}},
  \bibinfo {author} {\bibfnamefont {R.}~\bibnamefont {Yalovetzky}}, \bibinfo
  {author} {\bibfnamefont {P.}~\bibnamefont {Minssen}}, \bibinfo {author}
  {\bibfnamefont {D.}~\bibnamefont {Herman}}, \bibinfo {author} {\bibfnamefont
  {S.}~\bibnamefont {Hu}},\ and\ \bibinfo {author} {\bibfnamefont
  {M.}~\bibnamefont {Pistoia}},\ }\bibfield  {title} {\bibinfo {title}
  {{Constrained Quantum Optimization for Extractive Summarization on a
  Trapped-ion Quantum Computer}},\ }\href {http://arxiv.org/abs/2206.06290} {\
  (\bibinfo {year} {2022})},\ \Eprint {https://arxiv.org/abs/2206.06290}
  {arXiv:2206.06290} \BibitemShut {NoStop}%
\bibitem [{\citenamefont {Preskill}(2018)}]{Preskill2018}%
  \BibitemOpen
  \bibfield  {author} {\bibinfo {author} {\bibfnamefont {J.}~\bibnamefont
  {Preskill}},\ }\bibfield  {title} {\bibinfo {title} {{Quantum computing in
  the NISQ era and beyond}},\ }\href {https://doi.org/10.22331/q-2018-08-06-79}
  {\bibfield  {journal} {\bibinfo  {journal} {Quantum}\ }\textbf {\bibinfo
  {volume} {2}},\ \bibinfo {pages} {1} (\bibinfo {year} {2018})},\ \Eprint
  {https://arxiv.org/abs/1801.00862} {arXiv:1801.00862} \BibitemShut {NoStop}%
\bibitem [{\citenamefont {Shor}(1994)}]{shor1994factoring}%
  \BibitemOpen
  \bibfield  {author} {\bibinfo {author} {\bibfnamefont {P.~W.}\ \bibnamefont
  {Shor}},\ }\bibfield  {title} {\bibinfo {title} {Algorithms for quantum
  computation: discrete logarithms and factoring},\ }in\ \href
  {https://doi.org/10.1109/SFCS.1994.365700} {\emph {\bibinfo {booktitle}
  {Proceedings 35th Annual Symposium on Foundations of Computer Science}}}\
  (\bibinfo {year} {1994})\ pp.\ \bibinfo {pages} {124--134}\BibitemShut
  {NoStop}%
\bibitem [{\citenamefont {Shor}(1997)}]{shor1997algorithm}%
  \BibitemOpen
  \bibfield  {author} {\bibinfo {author} {\bibfnamefont {P.~W.}\ \bibnamefont
  {Shor}},\ }\bibfield  {title} {\bibinfo {title} {Polynomial-time algorithms
  for prime factorization and discrete logarithms on a quantum computer},\
  }\href {https://doi.org/10.1137/S0097539795293172} {\bibfield  {journal}
  {\bibinfo  {journal} {SIAM J. Comput.}\ }\textbf {\bibinfo {volume} {26}},\
  \bibinfo {pages} {1484} (\bibinfo {year} {1997})}\BibitemShut {NoStop}%
\bibitem [{\citenamefont {Cerezo}\ \emph {et~al.}(2021)\citenamefont {Cerezo},
  \citenamefont {Arrasmith}, \citenamefont {Babbush}, \citenamefont {Benjamin},
  \citenamefont {Endo}, \citenamefont {Fujii}, \citenamefont {Mcclean},
  \citenamefont {Mitarai}, \citenamefont {Yuan}, \citenamefont {Cincio},\ and\
  \citenamefont {Coles}}]{Cerezo2021}%
  \BibitemOpen
  \bibfield  {author} {\bibinfo {author} {\bibfnamefont {M.}~\bibnamefont
  {Cerezo}}, \bibinfo {author} {\bibfnamefont {A.}~\bibnamefont {Arrasmith}},
  \bibinfo {author} {\bibfnamefont {R.}~\bibnamefont {Babbush}}, \bibinfo
  {author} {\bibfnamefont {S.}~\bibnamefont {Benjamin}}, \bibinfo {author}
  {\bibfnamefont {S.}~\bibnamefont {Endo}}, \bibinfo {author} {\bibfnamefont
  {K.}~\bibnamefont {Fujii}}, \bibinfo {author} {\bibfnamefont
  {J.}~\bibnamefont {Mcclean}}, \bibinfo {author} {\bibfnamefont
  {K.}~\bibnamefont {Mitarai}}, \bibinfo {author} {\bibfnamefont
  {X.}~\bibnamefont {Yuan}}, \bibinfo {author} {\bibfnamefont {L.}~\bibnamefont
  {Cincio}},\ and\ \bibinfo {author} {\bibfnamefont {P.}~\bibnamefont
  {Coles}},\ }\bibfield  {title} {\bibinfo {title} {Variational quantum
  algorithms},\ }\href {https://doi.org/10.1038/s42254-021-00348-9} {\bibfield
  {journal} {\bibinfo  {journal} {Nature Reviews Physics}\ }\textbf {\bibinfo
  {volume} {3}},\ \bibinfo {pages} {1} (\bibinfo {year} {2021})}\BibitemShut
  {NoStop}%
\bibitem [{\citenamefont {Khairy}\ \emph {et~al.}(2020)\citenamefont {Khairy},
  \citenamefont {Shaydulin}, \citenamefont {Cincio}, \citenamefont {Alexeev},\
  and\ \citenamefont {Balaprakash}}]{Khairy2020}%
  \BibitemOpen
  \bibfield  {author} {\bibinfo {author} {\bibfnamefont {S.}~\bibnamefont
  {Khairy}}, \bibinfo {author} {\bibfnamefont {R.}~\bibnamefont {Shaydulin}},
  \bibinfo {author} {\bibfnamefont {L.}~\bibnamefont {Cincio}}, \bibinfo
  {author} {\bibfnamefont {Y.}~\bibnamefont {Alexeev}},\ and\ \bibinfo {author}
  {\bibfnamefont {P.}~\bibnamefont {Balaprakash}},\ }\bibfield  {title}
  {\bibinfo {title} {Learning to optimize variational quantum circuits to solve
  combinatorial problems},\ }\href {https://doi.org/10.1609/aaai.v34i03.5616}
  {\bibfield  {journal} {\bibinfo  {journal} {Proceedings of the AAAI
  Conference on Artificial Intelligence}\ }\textbf {\bibinfo {volume} {34}},\
  \bibinfo {pages} {2367} (\bibinfo {year} {2020})}\BibitemShut {NoStop}%
\bibitem [{\citenamefont {Apolloni}\ \emph {et~al.}(1989)\citenamefont
  {Apolloni}, \citenamefont {Carvalho},\ and\ \citenamefont
  {de~Falco}}]{Apolloni89}%
  \BibitemOpen
  \bibfield  {author} {\bibinfo {author} {\bibfnamefont {B.}~\bibnamefont
  {Apolloni}}, \bibinfo {author} {\bibfnamefont {C.}~\bibnamefont {Carvalho}},\
  and\ \bibinfo {author} {\bibfnamefont {D.}~\bibnamefont {de~Falco}},\
  }\bibfield  {title} {\bibinfo {title} {Quantum stochastic optimization},\
  }\href {https://doi.org/10.1016/0304-4149(89)90040-9} {\bibfield  {journal}
  {\bibinfo  {journal} {Stoch. Process. Their Appl.}\ }\textbf {\bibinfo
  {volume} {33}},\ \bibinfo {pages} {233 } (\bibinfo {year}
  {1989})}\BibitemShut {NoStop}%
\bibitem [{\citenamefont {Finnila}\ \emph {et~al.}(1994)\citenamefont
  {Finnila}, \citenamefont {Gomez}, \citenamefont {Sebenik}, \citenamefont
  {Stenson},\ and\ \citenamefont {Doll}}]{Finnila1994QuantumAnnealing}%
  \BibitemOpen
  \bibfield  {author} {\bibinfo {author} {\bibfnamefont {A.~B.}\ \bibnamefont
  {Finnila}}, \bibinfo {author} {\bibfnamefont {M.~A.}\ \bibnamefont {Gomez}},
  \bibinfo {author} {\bibfnamefont {C.}~\bibnamefont {Sebenik}}, \bibinfo
  {author} {\bibfnamefont {C.}~\bibnamefont {Stenson}},\ and\ \bibinfo {author}
  {\bibfnamefont {J.~D.}\ \bibnamefont {Doll}},\ }\bibfield  {title} {\bibinfo
  {title} {Quantum annealing: A new method for minimizing multidimensional
  functions},\ }\href {https://doi.org/10.1016/0009-2614(94)00117-0} {\bibfield
   {journal} {\bibinfo  {journal} {Chem. Phys. Lett.}\ }\textbf {\bibinfo
  {volume} {219}},\ \bibinfo {pages} {343 } (\bibinfo {year}
  {1994})}\BibitemShut {NoStop}%
\bibitem [{\citenamefont {Kadowaki}\ and\ \citenamefont
  {Nishimori}(1998{\natexlab{a}})}]{KadowakiNishimori1998QuantumAnnealing}%
  \BibitemOpen
  \bibfield  {author} {\bibinfo {author} {\bibfnamefont {T.}~\bibnamefont
  {Kadowaki}}\ and\ \bibinfo {author} {\bibfnamefont {H.}~\bibnamefont
  {Nishimori}},\ }\bibfield  {title} {\bibinfo {title} {Quantum annealing in
  the transverse ising model},\ }\href
  {https://doi.org/10.1103/PhysRevE.58.5355} {\bibfield  {journal} {\bibinfo
  {journal} {Phys. Rev. E}\ }\textbf {\bibinfo {volume} {58}},\ \bibinfo
  {pages} {5355} (\bibinfo {year} {1998}{\natexlab{a}})}\BibitemShut {NoStop}%
\bibitem [{\citenamefont {de~Falco}\ and\ \citenamefont
  {Tamascelli}(2011)}]{Falco2011}%
  \BibitemOpen
  \bibfield  {author} {\bibinfo {author} {\bibfnamefont {D.}~\bibnamefont
  {de~Falco}}\ and\ \bibinfo {author} {\bibfnamefont {D.}~\bibnamefont
  {Tamascelli}},\ }\bibfield  {title} {\bibinfo {title} {An introduction to
  quantum annealing},\ }\href {https://doi.org/10.1051/ita/2011013} {\bibfield
  {journal} {\bibinfo  {journal} {RAIRO - Theoretical Informatics and
  Applications}\ }\textbf {\bibinfo {volume} {45}} (\bibinfo {year}
  {2011})}\BibitemShut {NoStop}%
\bibitem [{\citenamefont {Ayanzadeh}\ \emph {et~al.}(2021)\citenamefont
  {Ayanzadeh}, \citenamefont {Dorband}, \citenamefont {Halem},\ and\
  \citenamefont {Finin}}]{Ayanzadeh2021}%
  \BibitemOpen
  \bibfield  {author} {\bibinfo {author} {\bibfnamefont {R.}~\bibnamefont
  {Ayanzadeh}}, \bibinfo {author} {\bibfnamefont {J.}~\bibnamefont {Dorband}},
  \bibinfo {author} {\bibfnamefont {M.}~\bibnamefont {Halem}},\ and\ \bibinfo
  {author} {\bibfnamefont {T.}~\bibnamefont {Finin}},\ }\bibfield  {title}
  {\bibinfo {title} {{Multi-qubit correction for quantum annealers}},\ }\href
  {https://doi.org/10.1038/s41598-021-95482-w} {\bibfield  {journal} {\bibinfo
  {journal} {Scientific Reports}\ }\textbf {\bibinfo {volume} {11}},\ \bibinfo
  {pages} {1} (\bibinfo {year} {2021})},\ \Eprint
  {https://arxiv.org/abs/2010.00115} {arXiv:2010.00115} \BibitemShut {NoStop}%
\bibitem [{\citenamefont {Willsch}\ \emph
  {et~al.}(2022{\natexlab{a}})\citenamefont {Willsch}, \citenamefont {Willsch},
  \citenamefont {Gonzalez~Calaza}, \citenamefont {Jin}, \citenamefont {{De
  Raedt}}, \citenamefont {Svensson},\ and\ \citenamefont
  {Michielsen}}]{Willsch2021BenchmarkAdvantage}%
  \BibitemOpen
  \bibfield  {author} {\bibinfo {author} {\bibfnamefont {D.}~\bibnamefont
  {Willsch}}, \bibinfo {author} {\bibfnamefont {M.}~\bibnamefont {Willsch}},
  \bibinfo {author} {\bibfnamefont {C.~D.}\ \bibnamefont {Gonzalez~Calaza}},
  \bibinfo {author} {\bibfnamefont {F.}~\bibnamefont {Jin}}, \bibinfo {author}
  {\bibfnamefont {H.}~\bibnamefont {{De Raedt}}}, \bibinfo {author}
  {\bibfnamefont {M.}~\bibnamefont {Svensson}},\ and\ \bibinfo {author}
  {\bibfnamefont {K.}~\bibnamefont {Michielsen}},\ }\bibfield  {title}
  {\bibinfo {title} {{B}enchmarking {A}dvantage and {D-Wave 2000Q} quantum
  annealers with exact cover problems},\ }\href
  {https://doi.org/10.1007/s11128-022-03476-y} {\bibfield  {journal} {\bibinfo
  {journal} {Quantum Inf. Process.}\ }\textbf {\bibinfo {volume} {21}},\
  \bibinfo {pages} {141} (\bibinfo {year} {2022}{\natexlab{a}})}\BibitemShut
  {NoStop}%
\bibitem [{\citenamefont {Heim}\ \emph {et~al.}(2015)\citenamefont {Heim},
  \citenamefont {R{\o}nnow}, \citenamefont {Isakov},\ and\ \citenamefont
  {Troyer}}]{Heim2015}%
  \BibitemOpen
  \bibfield  {author} {\bibinfo {author} {\bibfnamefont {B.}~\bibnamefont
  {Heim}}, \bibinfo {author} {\bibfnamefont {T.~F.}\ \bibnamefont {R{\o}nnow}},
  \bibinfo {author} {\bibfnamefont {S.~V.}\ \bibnamefont {Isakov}},\ and\
  \bibinfo {author} {\bibfnamefont {M.}~\bibnamefont {Troyer}},\ }\bibfield
  {title} {\bibinfo {title} {{Quantum versus classical annealing of Ising spin
  glasses}},\ }\href {https://doi.org/10.1126/science.aaa4170} {\bibfield
  {journal} {\bibinfo  {journal} {Science}\ }\textbf {\bibinfo {volume}
  {348}},\ \bibinfo {pages} {215} (\bibinfo {year} {2015})},\ \Eprint
  {https://arxiv.org/abs/1411.5693} {arXiv:1411.5693} \BibitemShut {NoStop}%
\bibitem [{\citenamefont {Yan}\ and\ \citenamefont {Sinitsyn}(2022)}]{Yan2022}%
  \BibitemOpen
  \bibfield  {author} {\bibinfo {author} {\bibfnamefont {B.}~\bibnamefont
  {Yan}}\ and\ \bibinfo {author} {\bibfnamefont {N.~A.}\ \bibnamefont
  {Sinitsyn}},\ }\bibfield  {title} {\bibinfo {title} {{Analytical solution for
  nonadiabatic quantum annealing to arbitrary Ising spin Hamiltonian}},\ }\href
  {https://doi.org/10.1038/s41467-022-29887-0} {\bibfield  {journal} {\bibinfo
  {journal} {Nature Communications}\ }\textbf {\bibinfo {volume} {13}},\
  \bibinfo {pages} {1} (\bibinfo {year} {2022})},\ \Eprint
  {https://arxiv.org/abs/2110.12354} {arXiv:2110.12354} \BibitemShut {NoStop}%
\bibitem [{\citenamefont {Tasseff}\ \emph {et~al.}(2022)\citenamefont
  {Tasseff}, \citenamefont {Albash}, \citenamefont {Morrell}, \citenamefont
  {Vuffray}, \citenamefont {Lokhov}, \citenamefont {Misra},\ and\ \citenamefont
  {Coffrin}}]{Tasseff2022}%
  \BibitemOpen
  \bibfield  {author} {\bibinfo {author} {\bibfnamefont {B.}~\bibnamefont
  {Tasseff}}, \bibinfo {author} {\bibfnamefont {T.}~\bibnamefont {Albash}},
  \bibinfo {author} {\bibfnamefont {Z.}~\bibnamefont {Morrell}}, \bibinfo
  {author} {\bibfnamefont {M.}~\bibnamefont {Vuffray}}, \bibinfo {author}
  {\bibfnamefont {A.~Y.}\ \bibnamefont {Lokhov}}, \bibinfo {author}
  {\bibfnamefont {S.}~\bibnamefont {Misra}},\ and\ \bibinfo {author}
  {\bibfnamefont {C.}~\bibnamefont {Coffrin}},\ }\bibfield  {title} {\bibinfo
  {title} {{On the Emerging Potential of Quantum Annealing Hardware for
  Combinatorial Optimization}},\ }\href {http://arxiv.org/abs/2210.04291} {\ ,\
  \bibinfo {pages} {1} (\bibinfo {year} {2022})},\ \Eprint
  {https://arxiv.org/abs/2210.04291} {arXiv:2210.04291} \BibitemShut {NoStop}%
\bibitem [{\citenamefont {Farhi}\ \emph {et~al.}(2014)\citenamefont {Farhi},
  \citenamefont {Goldstone},\ and\ \citenamefont {Gutmann}}]{Farhi2014}%
  \BibitemOpen
  \bibfield  {author} {\bibinfo {author} {\bibfnamefont {E.}~\bibnamefont
  {Farhi}}, \bibinfo {author} {\bibfnamefont {J.}~\bibnamefont {Goldstone}},\
  and\ \bibinfo {author} {\bibfnamefont {S.}~\bibnamefont {Gutmann}},\
  }\bibfield  {title} {\bibinfo {title} {A quantum approximate optimization
  algorithm},\ }\href@noop {} {\  (\bibinfo {year} {2014})}\BibitemShut
  {NoStop}%
\bibitem [{\citenamefont {Willsch}\ \emph {et~al.}(2020)\citenamefont
  {Willsch}, \citenamefont {Willsch}, \citenamefont {Jin}, \citenamefont {{De
  Raedt}},\ and\ \citenamefont {Michielsen}}]{Willsch2019BenchmarkingQAOA}%
  \BibitemOpen
  \bibfield  {author} {\bibinfo {author} {\bibfnamefont {M.}~\bibnamefont
  {Willsch}}, \bibinfo {author} {\bibfnamefont {D.}~\bibnamefont {Willsch}},
  \bibinfo {author} {\bibfnamefont {F.}~\bibnamefont {Jin}}, \bibinfo {author}
  {\bibfnamefont {H.}~\bibnamefont {{De Raedt}}},\ and\ \bibinfo {author}
  {\bibfnamefont {K.}~\bibnamefont {Michielsen}},\ }\bibfield  {title}
  {\bibinfo {title} {Benchmarking the quantum approximate optimization
  algorithm},\ }\href {https://doi.org/10.1007/s11128-020-02692-8} {\bibfield
  {journal} {\bibinfo  {journal} {Quantum Inf. Process.}\ }\textbf {\bibinfo
  {volume} {19}},\ \bibinfo {pages} {197} (\bibinfo {year} {2020})}\BibitemShut
  {NoStop}%
\bibitem [{\citenamefont {Jiang}\ \emph {et~al.}(2022)\citenamefont {Jiang},
  \citenamefont {Shen},\ and\ \citenamefont {Liu}}]{Jiang2022}%
  \BibitemOpen
  \bibfield  {author} {\bibinfo {author} {\bibfnamefont {H.}~\bibnamefont
  {Jiang}}, \bibinfo {author} {\bibfnamefont {Z.-J.~M.}\ \bibnamefont {Shen}},\
  and\ \bibinfo {author} {\bibfnamefont {J.}~\bibnamefont {Liu}},\ }\bibfield
  {title} {\bibinfo {title} {{Quantum Computing Methods for Supply Chain
  Management}},\ }\href {http://arxiv.org/abs/2209.08246} {\  (\bibinfo {year}
  {2022})},\ \Eprint {https://arxiv.org/abs/2209.08246} {arXiv:2209.08246}
  \BibitemShut {NoStop}%
\bibitem [{\citenamefont {Or{\'{u}}s}\ \emph {et~al.}(2019)\citenamefont
  {Or{\'{u}}s}, \citenamefont {Mugel},\ and\ \citenamefont
  {Lizaso}}]{Orus2019}%
  \BibitemOpen
  \bibfield  {author} {\bibinfo {author} {\bibfnamefont {R.}~\bibnamefont
  {Or{\'{u}}s}}, \bibinfo {author} {\bibfnamefont {S.}~\bibnamefont {Mugel}},\
  and\ \bibinfo {author} {\bibfnamefont {E.}~\bibnamefont {Lizaso}},\
  }\bibfield  {title} {\bibinfo {title} {{Quantum computing for finance:
  Overview and prospects}},\ }\href
  {https://doi.org/10.1016/j.revip.2019.100028} {\bibfield  {journal} {\bibinfo
   {journal} {Reviews in Physics}\ }\textbf {\bibinfo {volume} {4}},\ \bibinfo
  {pages} {1} (\bibinfo {year} {2019})},\ \Eprint
  {https://arxiv.org/abs/1807.03890} {arXiv:1807.03890} \BibitemShut {NoStop}%
\bibitem [{\citenamefont {Souza}\ \emph {et~al.}(2022)\citenamefont {Souza},
  \citenamefont {Martins}, \citenamefont {Roditi}, \citenamefont {S{\'{a}}},
  \citenamefont {Sarthour},\ and\ \citenamefont {Oliveira}}]{Souza2022}%
  \BibitemOpen
  \bibfield  {author} {\bibinfo {author} {\bibfnamefont {A.~M.}\ \bibnamefont
  {Souza}}, \bibinfo {author} {\bibfnamefont {E.~O.}\ \bibnamefont {Martins}},
  \bibinfo {author} {\bibfnamefont {I.}~\bibnamefont {Roditi}}, \bibinfo
  {author} {\bibfnamefont {N.}~\bibnamefont {S{\'{a}}}}, \bibinfo {author}
  {\bibfnamefont {R.~S.}\ \bibnamefont {Sarthour}},\ and\ \bibinfo {author}
  {\bibfnamefont {I.~S.}\ \bibnamefont {Oliveira}},\ }\bibfield  {title}
  {\bibinfo {title} {{An Application of Quantum Annealing Computing to Seismic
  Inversion}},\ }\href {https://doi.org/10.3389/fphy.2021.748285} {\bibfield
  {journal} {\bibinfo  {journal} {Frontiers in Physics}\ }\textbf {\bibinfo
  {volume} {9}},\ \bibinfo {pages} {1} (\bibinfo {year} {2022})},\ \Eprint
  {https://arxiv.org/abs/2005.02846} {arXiv:2005.02846} \BibitemShut {NoStop}%
\bibitem [{\citenamefont {Mugel}\ \emph {et~al.}(2022)\citenamefont {Mugel},
  \citenamefont {Kuchkovsky}, \citenamefont {S{\'{a}}nchez}, \citenamefont
  {Fern{\'{a}}ndez-Lorenzo}, \citenamefont {Luis-Hita}, \citenamefont
  {Lizaso},\ and\ \citenamefont {Or{\'{u}}s}}]{Mugel2022}%
  \BibitemOpen
  \bibfield  {author} {\bibinfo {author} {\bibfnamefont {S.}~\bibnamefont
  {Mugel}}, \bibinfo {author} {\bibfnamefont {C.}~\bibnamefont {Kuchkovsky}},
  \bibinfo {author} {\bibfnamefont {E.}~\bibnamefont {S{\'{a}}nchez}}, \bibinfo
  {author} {\bibfnamefont {S.}~\bibnamefont {Fern{\'{a}}ndez-Lorenzo}},
  \bibinfo {author} {\bibfnamefont {J.}~\bibnamefont {Luis-Hita}}, \bibinfo
  {author} {\bibfnamefont {E.}~\bibnamefont {Lizaso}},\ and\ \bibinfo {author}
  {\bibfnamefont {R.}~\bibnamefont {Or{\'{u}}s}},\ }\bibfield  {title}
  {\bibinfo {title} {{Dynamic portfolio optimization with real datasets using
  quantum processors and quantum-inspired tensor networks}},\ }\href
  {https://doi.org/10.1103/PhysRevResearch.4.013006} {\bibfield  {journal}
  {\bibinfo  {journal} {Physical Review Research}\ }\textbf {\bibinfo {volume}
  {4}},\ \bibinfo {pages} {1} (\bibinfo {year} {2022})},\ \Eprint
  {https://arxiv.org/abs/2007.00017} {arXiv:2007.00017} \BibitemShut {NoStop}%
\bibitem [{\citenamefont {Sharabiani}\ \emph {et~al.}(2021)\citenamefont
  {Sharabiani}, \citenamefont {Jakobsen}, \citenamefont {Jeppesen},\ and\
  \citenamefont {Mahani}}]{Sharabiani2021}%
  \BibitemOpen
  \bibfield  {author} {\bibinfo {author} {\bibfnamefont {M.~T.~A.}\
  \bibnamefont {Sharabiani}}, \bibinfo {author} {\bibfnamefont {V.~B.}\
  \bibnamefont {Jakobsen}}, \bibinfo {author} {\bibfnamefont {M.}~\bibnamefont
  {Jeppesen}},\ and\ \bibinfo {author} {\bibfnamefont {A.~S.}\ \bibnamefont
  {Mahani}},\ }\bibfield  {title} {\bibinfo {title} {{Quantum Computing in
  Green Energy Production}},\ }\href {http://arxiv.org/abs/2105.11322} {\ ,\
  \bibinfo {pages} {1} (\bibinfo {year} {2021})},\ \Eprint
  {https://arxiv.org/abs/2105.11322} {arXiv:2105.11322} \BibitemShut {NoStop}%
\bibitem [{\citenamefont {Urgelles}\ \emph {et~al.}(2022)\citenamefont
  {Urgelles}, \citenamefont {Picazo-martinez}, \citenamefont {Garcia-roger},\
  and\ \citenamefont {Monserrat}}]{Urgelles2022}%
  \BibitemOpen
  \bibfield  {author} {\bibinfo {author} {\bibfnamefont {H.}~\bibnamefont
  {Urgelles}}, \bibinfo {author} {\bibfnamefont {P.}~\bibnamefont
  {Picazo-martinez}}, \bibinfo {author} {\bibfnamefont {D.}~\bibnamefont
  {Garcia-roger}},\ and\ \bibinfo {author} {\bibfnamefont {J.~F.}\ \bibnamefont
  {Monserrat}},\ }\bibfield  {title} {\bibinfo {title} {{Multi-Objective
  Routing Optimization for 6G Communication Networks Using a Quantum
  Approximate Optimization Algorithm}},\ }\href@noop {} {\  (\bibinfo {year}
  {2022})}\BibitemShut {NoStop}%
\bibitem [{\citenamefont {Luckow}\ \emph {et~al.}(2021)\citenamefont {Luckow},
  \citenamefont {Klepsch},\ and\ \citenamefont {Pichlmeier}}]{Luckow2021}%
  \BibitemOpen
  \bibfield  {author} {\bibinfo {author} {\bibfnamefont {A.}~\bibnamefont
  {Luckow}}, \bibinfo {author} {\bibfnamefont {J.}~\bibnamefont {Klepsch}},\
  and\ \bibinfo {author} {\bibfnamefont {J.}~\bibnamefont {Pichlmeier}},\
  }\bibfield  {title} {\bibinfo {title} {{Quantum Computing: Towards Industry
  Reference Problems}},\ }\href {https://doi.org/10.1007/s42354-021-0335-7}
  {\bibfield  {journal} {\bibinfo  {journal} {Digitale Welt}\ }\textbf
  {\bibinfo {volume} {5}},\ \bibinfo {pages} {38} (\bibinfo {year} {2021})},\
  \Eprint {https://arxiv.org/abs/2103.07433} {arXiv:2103.07433} \BibitemShut
  {NoStop}%
\bibitem [{\citenamefont {Inoue}\ \emph {et~al.}(2021)\citenamefont {Inoue},
  \citenamefont {Okada}, \citenamefont {Matsumori}, \citenamefont {Aihara},\
  and\ \citenamefont {Yoshida}}]{Inoue2021}%
  \BibitemOpen
  \bibfield  {author} {\bibinfo {author} {\bibfnamefont {D.}~\bibnamefont
  {Inoue}}, \bibinfo {author} {\bibfnamefont {A.}~\bibnamefont {Okada}},
  \bibinfo {author} {\bibfnamefont {T.}~\bibnamefont {Matsumori}}, \bibinfo
  {author} {\bibfnamefont {K.}~\bibnamefont {Aihara}},\ and\ \bibinfo {author}
  {\bibfnamefont {H.}~\bibnamefont {Yoshida}},\ }\bibfield  {title} {\bibinfo
  {title} {{Traffic signal optimization on a square lattice with quantum
  annealing}},\ }\href {https://doi.org/10.1038/s41598-021-82740-0} {\bibfield
  {journal} {\bibinfo  {journal} {Scientific Reports}\ }\textbf {\bibinfo
  {volume} {11}},\ \bibinfo {pages} {1} (\bibinfo {year} {2021})},\ \Eprint
  {https://arxiv.org/abs/2003.07527} {arXiv:2003.07527} \BibitemShut {NoStop}%
\bibitem [{\citenamefont {{IBM Q team}}(2021)}]{Qiskit}%
  \BibitemOpen
  \bibfield  {author} {\bibinfo {author} {\bibnamefont {{IBM Q team}}},\ }\href
  {https://doi.org/10.5281/zenodo.2573505} {\bibinfo {title} {Qiskit: An
  open-source framework for quantum computing}} (\bibinfo {year} {2021}),\
  \bibinfo {note} {\url{https://doi.org/10.5281/zenodo.2573505}, release
  0.39.4}\BibitemShut {NoStop}%
\bibitem [{\citenamefont {{D-Wave Systems}}(2022)}]{DWOceanSDK}%
  \BibitemOpen
  \bibfield  {author} {\bibinfo {author} {\bibnamefont {{D-Wave Systems}}},\
  }\href {https://github.com/dwavesystems/dwave-ocean-sdk} {\bibinfo {title}
  {{D-Wave Ocean SDK}}} (\bibinfo {year} {2022}),\ \bibinfo {note}
  {\url{https://github.com/dwavesystems/dwave-ocean-sdk}, release
  6.2.0}\BibitemShut {NoStop}%
\bibitem [{\citenamefont {Glover}\ \emph {et~al.}(2019)\citenamefont {Glover},
  \citenamefont {Kochenberger},\ and\ \citenamefont {Du}}]{Glover2019}%
  \BibitemOpen
  \bibfield  {author} {\bibinfo {author} {\bibfnamefont {F.}~\bibnamefont
  {Glover}}, \bibinfo {author} {\bibfnamefont {G.}~\bibnamefont
  {Kochenberger}},\ and\ \bibinfo {author} {\bibfnamefont {Y.}~\bibnamefont
  {Du}},\ }\bibfield  {title} {\bibinfo {title} {{Quantum Bridge Analytics I: a
  tutorial on formulating and using QUBO models}},\ }\href
  {https://doi.org/10.1007/s10288-019-00424-y} {\bibfield  {journal} {\bibinfo
  {journal} {4or}\ }\textbf {\bibinfo {volume} {17}},\ \bibinfo {pages} {335}
  (\bibinfo {year} {2019})}\BibitemShut {NoStop}%
\bibitem [{\citenamefont {Glover}\ \emph {et~al.}(2022)\citenamefont {Glover},
  \citenamefont {Kochenberger}, \citenamefont {Ma},\ and\ \citenamefont
  {Du}}]{Glover2022}%
  \BibitemOpen
  \bibfield  {author} {\bibinfo {author} {\bibfnamefont {F.}~\bibnamefont
  {Glover}}, \bibinfo {author} {\bibfnamefont {G.}~\bibnamefont
  {Kochenberger}}, \bibinfo {author} {\bibfnamefont {M.}~\bibnamefont {Ma}},\
  and\ \bibinfo {author} {\bibfnamefont {Y.}~\bibnamefont {Du}},\ }\bibfield
  {title} {\bibinfo {title} {{Quantum Bridge Analytics II: QUBO-Plus, network
  optimization and combinatorial chaining for asset exchange}},\ }\href
  {https://doi.org/10.1007/s10479-022-04695-3} {\bibfield  {journal} {\bibinfo
  {journal} {Annals of Operations Research}\ }\textbf {\bibinfo {volume}
  {314}},\ \bibinfo {pages} {185} (\bibinfo {year} {2022})}\BibitemShut
  {NoStop}%
\bibitem [{\citenamefont {Sharma}\ \emph {et~al.}(2022)\citenamefont {Sharma},
  \citenamefont {Saharan}, \citenamefont {Chiew}, \citenamefont {Chiacchio},
  \citenamefont {Disilvestro}, \citenamefont {Demarie},\ and\ \citenamefont
  {Munro}}]{Sharma2022}%
  \BibitemOpen
  \bibfield  {author} {\bibinfo {author} {\bibfnamefont {V.}~\bibnamefont
  {Sharma}}, \bibinfo {author} {\bibfnamefont {N.~S.~B.}\ \bibnamefont
  {Saharan}}, \bibinfo {author} {\bibfnamefont {S.-H.}\ \bibnamefont {Chiew}},
  \bibinfo {author} {\bibfnamefont {E.~I.~R.}\ \bibnamefont {Chiacchio}},
  \bibinfo {author} {\bibfnamefont {L.}~\bibnamefont {Disilvestro}}, \bibinfo
  {author} {\bibfnamefont {T.~F.}\ \bibnamefont {Demarie}},\ and\ \bibinfo
  {author} {\bibfnamefont {E.}~\bibnamefont {Munro}},\ }\bibfield  {title}
  {\bibinfo {title} {{OpenQAOA -- An SDK for QAOA}},\ }\href
  {http://arxiv.org/abs/2210.08695} {\ ,\ \bibinfo {pages} {1} (\bibinfo {year}
  {2022})},\ \Eprint {https://arxiv.org/abs/2210.08695} {arXiv:2210.08695}
  \BibitemShut {NoStop}%
\bibitem [{\citenamefont {{De Raedt}}\ \emph {et~al.}(2007)\citenamefont {{De
  Raedt}}, \citenamefont {Michielsen}, \citenamefont {{De Raedt}},
  \citenamefont {Trieu}, \citenamefont {Arnold}, \citenamefont {Richter},
  \citenamefont {Lippert}, \citenamefont {Watanabe},\ and\ \citenamefont
  {Ito}}]{DeRaedt2007MassivelyParallel}%
  \BibitemOpen
  \bibfield  {author} {\bibinfo {author} {\bibfnamefont {K.}~\bibnamefont {{De
  Raedt}}}, \bibinfo {author} {\bibfnamefont {K.}~\bibnamefont {Michielsen}},
  \bibinfo {author} {\bibfnamefont {H.}~\bibnamefont {{De Raedt}}}, \bibinfo
  {author} {\bibfnamefont {B.}~\bibnamefont {Trieu}}, \bibinfo {author}
  {\bibfnamefont {G.}~\bibnamefont {Arnold}}, \bibinfo {author} {\bibfnamefont
  {M.}~\bibnamefont {Richter}}, \bibinfo {author} {\bibfnamefont {{\relax
  Th}.}~\bibnamefont {Lippert}}, \bibinfo {author} {\bibfnamefont
  {H.}~\bibnamefont {Watanabe}},\ and\ \bibinfo {author} {\bibfnamefont
  {N.}~\bibnamefont {Ito}},\ }\bibfield  {title} {\bibinfo {title} {Massively
  parallel quantum computer simulator},\ }\href
  {https://doi.org/10.1016/j.cpc.2006.08.007} {\bibfield  {journal} {\bibinfo
  {journal} {Comput. Phys. Commun.}\ }\textbf {\bibinfo {volume} {176}},\
  \bibinfo {pages} {121} (\bibinfo {year} {2007})}\BibitemShut {NoStop}%
\bibitem [{\citenamefont {{De Raedt}}\ \emph {et~al.}(2019)\citenamefont {{De
  Raedt}}, \citenamefont {Jin}, \citenamefont {Willsch}, \citenamefont
  {Willsch}, \citenamefont {Yoshioka}, \citenamefont {Ito}, \citenamefont
  {Yuan},\ and\ \citenamefont {Michielsen}}]{DeRaedt2018MassivelyParallel}%
  \BibitemOpen
  \bibfield  {author} {\bibinfo {author} {\bibfnamefont {H.}~\bibnamefont {{De
  Raedt}}}, \bibinfo {author} {\bibfnamefont {F.}~\bibnamefont {Jin}}, \bibinfo
  {author} {\bibfnamefont {D.}~\bibnamefont {Willsch}}, \bibinfo {author}
  {\bibfnamefont {M.}~\bibnamefont {Willsch}}, \bibinfo {author} {\bibfnamefont
  {N.}~\bibnamefont {Yoshioka}}, \bibinfo {author} {\bibfnamefont
  {N.}~\bibnamefont {Ito}}, \bibinfo {author} {\bibfnamefont {S.}~\bibnamefont
  {Yuan}},\ and\ \bibinfo {author} {\bibfnamefont {K.}~\bibnamefont
  {Michielsen}},\ }\bibfield  {title} {\bibinfo {title} {Massively parallel
  quantum computer simulator, eleven years later},\ }\href
  {https://doi.org/10.1016/j.cpc.2018.11.005} {\bibfield  {journal} {\bibinfo
  {journal} {Comput. Phys. Commun.}\ }\textbf {\bibinfo {volume} {237}},\
  \bibinfo {pages} {47 } (\bibinfo {year} {2019})}\BibitemShut {NoStop}%
\bibitem [{\citenamefont {Willsch}\ \emph
  {et~al.}(2022{\natexlab{b}})\citenamefont {Willsch}, \citenamefont {Willsch},
  \citenamefont {Jin}, \citenamefont {Michielsen},\ and\ \citenamefont {{De
  Raedt}}}]{Willsch2021JUQCSGQAOA}%
  \BibitemOpen
  \bibfield  {author} {\bibinfo {author} {\bibfnamefont {D.}~\bibnamefont
  {Willsch}}, \bibinfo {author} {\bibfnamefont {M.}~\bibnamefont {Willsch}},
  \bibinfo {author} {\bibfnamefont {F.}~\bibnamefont {Jin}}, \bibinfo {author}
  {\bibfnamefont {K.}~\bibnamefont {Michielsen}},\ and\ \bibinfo {author}
  {\bibfnamefont {H.}~\bibnamefont {{De Raedt}}},\ }\bibfield  {title}
  {\bibinfo {title} {{GPU}-accelerated simulations of quantum annealing and the
  quantum approximate optimization algorithm},\ }\href
  {https://doi.org/10.1016/j.cpc.2022.108411} {\bibfield  {journal} {\bibinfo
  {journal} {Comput. Phys. Commun.}\ }\textbf {\bibinfo {volume} {278}},\
  \bibinfo {pages} {108411} (\bibinfo {year} {2022}{\natexlab{b}})}\BibitemShut
  {NoStop}%
\bibitem [{Ver(2022)}]{Verma2022}%
  \BibitemOpen
  \bibfield  {title} {\bibinfo {title} {{Penalty and partitioning techniques to
  improve performance of QUBO solvers}},\ }\href
  {https://doi.org/10.1016/j.disopt.2020.100594} {\bibfield  {journal}
  {\bibinfo  {journal} {Discrete Optimization}\ }\textbf {\bibinfo {volume}
  {44}},\ \bibinfo {pages} {100594} (\bibinfo {year} {2022})}\BibitemShut
  {NoStop}%
\bibitem [{\citenamefont {Helsgaun}(2006)}]{Helsgaun2006}%
  \BibitemOpen
  \bibfield  {author} {\bibinfo {author} {\bibfnamefont {K.}~\bibnamefont
  {Helsgaun}},\ }\bibfield  {title} {\bibinfo {title} {{An Effective
  Implementation of K-opt Moves for the Lin-Kernighan TSP Heuristic}},\
  }\href@noop {} {\bibfield  {journal} {\bibinfo  {journal} {Writings on
  Computer Science}\ ,\ \bibinfo {pages} {1}} (\bibinfo {year}
  {2006})}\BibitemShut {NoStop}%
\bibitem [{\citenamefont {Applegate}\ \emph {et~al.}(2006)\citenamefont
  {Applegate}, \citenamefont {Bixby}, \citenamefont {Chvatál},\ and\
  \citenamefont {Cook}}]{Applegate2006}%
  \BibitemOpen
  \bibfield  {author} {\bibinfo {author} {\bibfnamefont {D.~L.}\ \bibnamefont
  {Applegate}}, \bibinfo {author} {\bibfnamefont {R.~E.}\ \bibnamefont
  {Bixby}}, \bibinfo {author} {\bibfnamefont {V.}~\bibnamefont {Chvatál}},\
  and\ \bibinfo {author} {\bibfnamefont {W.~J.}\ \bibnamefont {Cook}},\ }\href
  {http://www.jstor.org/stable/j.ctt7s8xg} {\emph {\bibinfo {title} {The
  Traveling Salesman Problem: A Computational Study}}}\ (\bibinfo  {publisher}
  {Princeton University Press},\ \bibinfo {year} {2006})\BibitemShut {NoStop}%
\bibitem [{\citenamefont {Applegate}\ \emph {et~al.}(1997)\citenamefont
  {Applegate}, \citenamefont {Bixby}, \citenamefont {Chvátal},\ and\
  \citenamefont {Cook}}]{concorde-website}%
  \BibitemOpen
  \bibfield  {author} {\bibinfo {author} {\bibfnamefont {D.~L.}\ \bibnamefont
  {Applegate}}, \bibinfo {author} {\bibfnamefont {R.~E.}\ \bibnamefont
  {Bixby}}, \bibinfo {author} {\bibfnamefont {V.}~\bibnamefont {Chvátal}},\
  and\ \bibinfo {author} {\bibfnamefont {W.~J.}\ \bibnamefont {Cook}},\
  }\href@noop {} {\bibinfo {title} {Concorde tsp solver}},\ \bibinfo
  {howpublished} {\url{https://www.math.uwaterloo.ca/tsp/concorde.html}}
  (\bibinfo {year} {1997})\BibitemShut {NoStop}%
\bibitem [{TSP(1995)}]{TSPLIB}%
  \BibitemOpen
  \href@noop {} {\bibinfo {title} {{TSPLIB95}: {A} library of traveling
  salesman and related problems}},\ \bibinfo {howpublished}
  {\url{http://comopt.ifi.uni-heidelberg.de/software/TSPLIB95/}} (\bibinfo
  {year} {1995})\BibitemShut {NoStop}%
\bibitem [{\citenamefont {Gr{\"{o}}tschel}\ and\ \citenamefont
  {Nemhauser}(2008)}]{Grotschel2008}%
  \BibitemOpen
  \bibfield  {author} {\bibinfo {author} {\bibfnamefont {M.}~\bibnamefont
  {Gr{\"{o}}tschel}}\ and\ \bibinfo {author} {\bibfnamefont {G.~L.}\
  \bibnamefont {Nemhauser}},\ }\bibfield  {title} {\bibinfo {title} {{George
  Dantzig's contributions to integer programming}},\ }\href
  {https://doi.org/10.1016/j.disopt.2007.08.003} {\bibfield  {journal}
  {\bibinfo  {journal} {Discrete Optimization}\ }\textbf {\bibinfo {volume}
  {5}},\ \bibinfo {pages} {168} (\bibinfo {year} {2008})}\BibitemShut {NoStop}%
\bibitem [{\citenamefont {Martello}\ and\ \citenamefont
  {Toth}(1990)}]{Martello1990}%
  \BibitemOpen
  \bibfield  {author} {\bibinfo {author} {\bibfnamefont {S.}~\bibnamefont
  {Martello}}\ and\ \bibinfo {author} {\bibfnamefont {P.}~\bibnamefont
  {Toth}},\ }\href@noop {} {\emph {\bibinfo {title} {Knapsack problems:
  algorithms and computer implementations}}}\ (\bibinfo  {publisher} {John
  Wiley \& Sons, Inc.},\ \bibinfo {address} {USA},\ \bibinfo {year}
  {1990})\BibitemShut {NoStop}%
\bibitem [{\citenamefont {He{\ss}ler}\ \emph {et~al.}(2020)\citenamefont
  {He{\ss}ler}, \citenamefont {Irnich}, \citenamefont {Kreiter},\ and\
  \citenamefont {Pferschy}}]{Hessler2020}%
  \BibitemOpen
  \bibfield  {author} {\bibinfo {author} {\bibfnamefont {K.}~\bibnamefont
  {He{\ss}ler}}, \bibinfo {author} {\bibfnamefont {S.}~\bibnamefont {Irnich}},
  \bibinfo {author} {\bibfnamefont {T.}~\bibnamefont {Kreiter}},\ and\ \bibinfo
  {author} {\bibfnamefont {U.}~\bibnamefont {Pferschy}},\ }\bibfield  {title}
  {\bibinfo {title} {{Lexicographic Bin-Packing Optimization for Loading Trucks
  in a Direct-Shipping System}},\ }\href
  {https://ideas.repec.org/p/jgu/wpaper/2009.html} {\  (\bibinfo {year}
  {2020})}\BibitemShut {NoStop}%
\bibitem [{\citenamefont {Yan}\ \emph {et~al.}(2022)\citenamefont {Yan},
  \citenamefont {Lu}, \citenamefont {Chen}, \citenamefont {Qin}, \citenamefont
  {Fang}, \citenamefont {Lin}, \citenamefont {Moscibroda}, \citenamefont
  {Rajmohan},\ and\ \citenamefont {Zhang}}]{Yan2022Book}%
  \BibitemOpen
  \bibfield  {author} {\bibinfo {author} {\bibfnamefont {J.}~\bibnamefont
  {Yan}}, \bibinfo {author} {\bibfnamefont {Y.}~\bibnamefont {Lu}}, \bibinfo
  {author} {\bibfnamefont {L.}~\bibnamefont {Chen}}, \bibinfo {author}
  {\bibfnamefont {S.}~\bibnamefont {Qin}}, \bibinfo {author} {\bibfnamefont
  {Y.}~\bibnamefont {Fang}}, \bibinfo {author} {\bibfnamefont {Q.}~\bibnamefont
  {Lin}}, \bibinfo {author} {\bibfnamefont {T.}~\bibnamefont {Moscibroda}},
  \bibinfo {author} {\bibfnamefont {S.}~\bibnamefont {Rajmohan}},\ and\
  \bibinfo {author} {\bibfnamefont {D.}~\bibnamefont {Zhang}},\ }\href
  {https://doi.org/10.1145/3534678.3539334} {\emph {\bibinfo {title}
  {Proceedings of the 28th ACM SIGKDD Conference on Knowledge Discovery and
  Data Mining, 2022, Washington, DC, USA}}},\ Vol.~\bibinfo {volume} {1}\
  (\bibinfo  {publisher} {Association for Computing Machinery},\ \bibinfo
  {year} {2022})\ \Eprint {https://arxiv.org/abs/2207.11122} {arXiv:2207.11122}
  \BibitemShut {NoStop}%
\bibitem [{\citenamefont {Kroes}\ \emph {et~al.}(2020)\citenamefont {Kroes},
  \citenamefont {Petrica}, \citenamefont {Cotofana},\ and\ \citenamefont
  {Blott}}]{Kroes2020}%
  \BibitemOpen
  \bibfield  {author} {\bibinfo {author} {\bibfnamefont {M.}~\bibnamefont
  {Kroes}}, \bibinfo {author} {\bibfnamefont {L.}~\bibnamefont {Petrica}},
  \bibinfo {author} {\bibfnamefont {S.}~\bibnamefont {Cotofana}},\ and\
  \bibinfo {author} {\bibfnamefont {M.}~\bibnamefont {Blott}},\ }\bibfield
  {title} {\bibinfo {title} {{Evolutionary bin packing for memory-efficient
  dataflow inference acceleration on FPGA}},\ }\href
  {https://doi.org/10.1145/3377930.3389808} {\bibfield  {journal} {\bibinfo
  {journal} {GECCO 2020 - Proceedings of the 2020 Genetic and Evolutionary
  Computation Conference}\ ,\ \bibinfo {pages} {1125}} (\bibinfo {year}
  {2020})},\ \Eprint {https://arxiv.org/abs/2003.12449} {arXiv:2003.12449}
  \BibitemShut {NoStop}%
\bibitem [{\citenamefont {Cheeseman}\ \emph {et~al.}(1991)\citenamefont
  {Cheeseman}, \citenamefont {Kanefsky},\ and\ \citenamefont
  {Taylor}}]{Cheeseman1991}%
  \BibitemOpen
  \bibfield  {author} {\bibinfo {author} {\bibfnamefont {P.}~\bibnamefont
  {Cheeseman}}, \bibinfo {author} {\bibfnamefont {B.}~\bibnamefont
  {Kanefsky}},\ and\ \bibinfo {author} {\bibfnamefont {W.~M.}\ \bibnamefont
  {Taylor}},\ }\bibfield  {title} {\bibinfo {title} {{Where the Really Hard
  Problems Are}},\ }\href@noop {} {\bibfield  {journal} {\bibinfo  {journal}
  {The 12nd International Joint Conference on Artificial Intelligence}\ ,\
  \bibinfo {pages} {331}} (\bibinfo {year} {1991})}\BibitemShut {NoStop}%
\bibitem [{\citenamefont {Kadowaki}\ and\ \citenamefont
  {Nishimori}(1998{\natexlab{b}})}]{Kadowaki1998}%
  \BibitemOpen
  \bibfield  {author} {\bibinfo {author} {\bibfnamefont {T.}~\bibnamefont
  {Kadowaki}}\ and\ \bibinfo {author} {\bibfnamefont {H.}~\bibnamefont
  {Nishimori}},\ }\bibfield  {title} {\bibinfo {title} {Quantum annealing in
  the transverse ising model},\ }\href@noop {} {\bibfield  {journal} {\bibinfo
  {journal} {Physical Review E}\ }\textbf {\bibinfo {volume} {58}} (\bibinfo
  {year} {1998}{\natexlab{b}})}\BibitemShut {NoStop}%
\bibitem [{\citenamefont {Johnson}\ \emph {et~al.}(2011)\citenamefont
  {Johnson}, \citenamefont {Amin}, \citenamefont {Gildert},\ and\ \citenamefont
  {et~al.}}]{Johnson2011}%
  \BibitemOpen
  \bibfield  {author} {\bibinfo {author} {\bibfnamefont {M.}~\bibnamefont
  {Johnson}}, \bibinfo {author} {\bibfnamefont {M.}~\bibnamefont {Amin}},
  \bibinfo {author} {\bibfnamefont {S.}~\bibnamefont {Gildert}},\ and\ \bibinfo
  {author} {\bibnamefont {et~al.}},\ }\bibfield  {title} {\bibinfo {title}
  {Quantum annealing with manufactured spins},\ }\href
  {https://doi.org/10.1038/nature10012} {\bibfield  {journal} {\bibinfo
  {journal} {Nature}\ }\textbf {\bibinfo {volume} {473}},\ \bibinfo {pages}
  {194} (\bibinfo {year} {2011})}\BibitemShut {NoStop}%
\bibitem [{\citenamefont {Wang}\ \emph {et~al.}(2019)\citenamefont {Wang},
  \citenamefont {Rubin}, \citenamefont {Dominy},\ and\ \citenamefont
  {Rieffel}}]{Wang2019}%
  \BibitemOpen
  \bibfield  {author} {\bibinfo {author} {\bibfnamefont {Z.}~\bibnamefont
  {Wang}}, \bibinfo {author} {\bibfnamefont {N.~C.}\ \bibnamefont {Rubin}},
  \bibinfo {author} {\bibfnamefont {J.~M.}\ \bibnamefont {Dominy}},\ and\
  \bibinfo {author} {\bibfnamefont {E.~G.}\ \bibnamefont {Rieffel}},\
  }\bibfield  {title} {\bibinfo {title} {{$XY$-mixers: analytical and numerical
  results for QAOA}}\ }\href {https://doi.org/10.1103/PhysRevA.101.012320}
  {10.1103/PhysRevA.101.012320} (\bibinfo {year} {2019}),\ \Eprint
  {https://arxiv.org/abs/1904.09314} {arXiv:1904.09314} \BibitemShut {NoStop}%
\bibitem [{\citenamefont {Willsch}\ \emph {et~al.}()\citenamefont {Willsch},
  \citenamefont {Jattana}, \citenamefont {Willsch}, \citenamefont {Schulz},
  \citenamefont {Jin}, \citenamefont {{De Raedt}},\ and\ \citenamefont
  {Michielsen}}]{Willsch2022HybridQuantumClassicalSimulations}%
  \BibitemOpen
  \bibfield  {author} {\bibinfo {author} {\bibfnamefont {D.}~\bibnamefont
  {Willsch}}, \bibinfo {author} {\bibfnamefont {M.}~\bibnamefont {Jattana}},
  \bibinfo {author} {\bibfnamefont {M.}~\bibnamefont {Willsch}}, \bibinfo
  {author} {\bibfnamefont {S.}~\bibnamefont {Schulz}}, \bibinfo {author}
  {\bibfnamefont {F.}~\bibnamefont {Jin}}, \bibinfo {author} {\bibfnamefont
  {H.}~\bibnamefont {{De Raedt}}},\ and\ \bibinfo {author} {\bibfnamefont
  {K.}~\bibnamefont {Michielsen}},\ }\bibfield  {title} {\bibinfo {title}
  {{H}ybrid {Q}uantum {C}lassical {S}imulations},\ }in\ \href
  {https://arxiv.org/abs/2210.02811} {\emph {\bibinfo {booktitle} {{NIC}
  {S}ymposium 2022}}},\ \bibinfo {series} {Publication Series of the John von
  Neumann Institute for Computing (NIC) NIC Series}, Vol.~\bibinfo {volume}
  {51},\ \bibinfo {editor} {edited by\ \bibinfo {editor} {\bibfnamefont
  {M.}~\bibnamefont {M\"uller}}, \bibinfo {editor} {\bibfnamefont
  {C.}~\bibnamefont {Peter}},\ and\ \bibinfo {editor} {\bibfnamefont
  {A.}~\bibnamefont {Trautmann}}}\ (\bibinfo  {publisher} {Forschungszentrum
  J\"ulich GmbH Zentralbibliothek, Verlag},\ \bibinfo {address} {J\"ulich})\
  pp.\ \bibinfo {pages} {31--44}\BibitemShut {NoStop}%
\bibitem [{\citenamefont {Kesselheim}\ \emph {et~al.}(2021)\citenamefont
  {Kesselheim}, \citenamefont {Herten}, \citenamefont {Krajsek}, \citenamefont
  {Ebert}, \citenamefont {Jitsev}, \citenamefont {Cherti}, \citenamefont
  {Langguth}, \citenamefont {Gong}, \citenamefont {Stadtler}, \citenamefont
  {Mozaffari}, \citenamefont {Cavallaro}, \citenamefont {Sedona}, \citenamefont
  {Schug}, \citenamefont {Strube}, \citenamefont {Kamath}, \citenamefont
  {Schultz}, \citenamefont {Riedel},\ and\ \citenamefont
  {Lippert}}]{JUWELSBooster}%
  \BibitemOpen
  \bibfield  {author} {\bibinfo {author} {\bibfnamefont {S.}~\bibnamefont
  {Kesselheim}}, \bibinfo {author} {\bibfnamefont {A.}~\bibnamefont {Herten}},
  \bibinfo {author} {\bibfnamefont {K.}~\bibnamefont {Krajsek}}, \bibinfo
  {author} {\bibfnamefont {J.}~\bibnamefont {Ebert}}, \bibinfo {author}
  {\bibfnamefont {J.}~\bibnamefont {Jitsev}}, \bibinfo {author} {\bibfnamefont
  {M.}~\bibnamefont {Cherti}}, \bibinfo {author} {\bibfnamefont
  {M.}~\bibnamefont {Langguth}}, \bibinfo {author} {\bibfnamefont
  {B.}~\bibnamefont {Gong}}, \bibinfo {author} {\bibfnamefont {S.}~\bibnamefont
  {Stadtler}}, \bibinfo {author} {\bibfnamefont {A.}~\bibnamefont {Mozaffari}},
  \bibinfo {author} {\bibfnamefont {G.}~\bibnamefont {Cavallaro}}, \bibinfo
  {author} {\bibfnamefont {R.}~\bibnamefont {Sedona}}, \bibinfo {author}
  {\bibfnamefont {A.}~\bibnamefont {Schug}}, \bibinfo {author} {\bibfnamefont
  {A.}~\bibnamefont {Strube}}, \bibinfo {author} {\bibfnamefont
  {R.}~\bibnamefont {Kamath}}, \bibinfo {author} {\bibfnamefont {M.~G.}\
  \bibnamefont {Schultz}}, \bibinfo {author} {\bibfnamefont {M.}~\bibnamefont
  {Riedel}},\ and\ \bibinfo {author} {\bibfnamefont {{\relax Th}.}~\bibnamefont
  {Lippert}},\ }\bibfield  {title} {\bibinfo {title} {{JUWELS Booster -- A
  Supercomputer for Large-Scale AI Research}},\ }\Eprint
  {https://arxiv.org/abs/2108.11976} {arXiv:2108.11976 [cs.DC]}  (\bibinfo
  {year} {2021})\BibitemShut {NoStop}%
\bibitem [{\citenamefont {Montanez-Barrera}\ \emph {et~al.}(2023)\citenamefont
  {Montanez-Barrera}, \citenamefont {van~den Heuvel}, \citenamefont {Willsch},\
  and\ \citenamefont {Michielsen}}]{Montanez-Barrera2023}%
  \BibitemOpen
  \bibfield  {author} {\bibinfo {author} {\bibfnamefont {J.~A.}\ \bibnamefont
  {Montanez-Barrera}}, \bibinfo {author} {\bibfnamefont {P.}~\bibnamefont
  {van~den Heuvel}}, \bibinfo {author} {\bibfnamefont {D.}~\bibnamefont
  {Willsch}},\ and\ \bibinfo {author} {\bibfnamefont {K.}~\bibnamefont
  {Michielsen}},\ }\bibfield  {title} {\bibinfo {title} {{Improving Performance
  in Combinatorial Optimization Problems with Inequality Constraints: An
  Evaluation of the Unbalanced Penalization Method on D-Wave Advantage}},\
  }\href {http://arxiv.org/abs/2305.18757} {\  (\bibinfo {year} {2023})},\
  \Eprint {https://arxiv.org/abs/2305.18757} {arXiv:2305.18757} \BibitemShut
  {NoStop}%
\end{thebibliography}%

\end{document}